# Primordial origins of Earth's carbon

**Bernard Marty**

*Centre de Recherches Pétrographiques et Géochimiques, CNRS,*
*Université de Lorraine, BP 20,*
*54220 Vandoeuvre les Nancy Cedex, France*

bmarty@crpg.cnrs-nancy.fr

**Conel M. O'D. Alexander**

*Department of Terrestrial Magnetism*
*Carnegie Institution of Washington*
*5241 Broad Branch Road, NW*
*Washington, DC 20015-1305, USA*

alexander@dtm.ciw.edu

**Sean N. Raymond**

*Univ. Bordeaux and CNRS*
*Laboratoire d'Astrophysique de Bordeaux, UMR 5804,*
*F-33270, Floirac, France.*
rayray.sean@gmail.com

# INTRODUCTION

It is commonly assumed that the building blocks of the terrestrial planets were derived from a cosmochemical reservoir that is best represented by chondrites, the so-called chondritic Earth model. This view is possibly a good approximation for refractory elements (although it has been recently questioned, e.g., Caro et al. 2008), but for volatile elements, other cosmochemical reservoirs might have contributed to the Earth, such as the solar nebula gas and/or cometary matter (Dauphas, 2003; Owen et al. 1992; Pepin, 2006). Hence, in order to get insights into the origin of the carbon in the Earth, it is necessary to compare: (i) the elemental abundances and isotopic compositions of not only carbon, as well as other volatile elements in potential cosmochemical "ancestors", and (ii) these compositions with those of terrestrial volatiles. This approach is the only one that has the potential for understanding the origin of the carbon in the Earth but, nevertheless, it has several intrinsic limitations. First, the terrestrial carbon budget is not well known, and, for the deep reservoir(s) such as the core and the lower mantle, is highly model-dependent. Second, the cosmochemical reservoir(s) that contributed volatile elements to the proto-Earth may not exist anymore because planet formation might have completely exhausted them (most of the mass present in the inner solar system is now in Venus and Earth). Third, planetary formation processes (accretion, differentiation, early evolution of the atmospheres) might have drastically modified the original elemental and isotopic compositions of the volatile elements in the Earth. Despite these limitations, robust constraints on the origin(s) of the carbon in the Earth can be deduced from comparative planetology of volatile elements, which are the focus of this chapter.

Carbon in the cloud of gas and dust from which the solar system formed was probably mainly in the forms of gaseous CO and of organic-rich carbonaceous dust (e.g., Zubko et al, 2004). Other forms of carbon, such as $CH_4$, volatile organics and diamonds, were also probably present but in much smaller proportions. In primitive meteorites, the depletion of carbon relative to refractory elements and normalized to the solar composition indicates that indeed a large fraction of this element was not in refractory phase(s) and had a volatile element behaviour like hydrogen, nitrogen and noble gases (e.g., Anders and Grevesse, 1989; Lodders, 2003). Furthermore, in primitive meteorites, the total amounts of carbon, nitrogen and trapped noble gases tend to correlate (Otting and Zärhinger, 1967; Kerridge, 1985). Thus, to first order the noble gases can be taken as proxies for the behaviour of carbon during planetary building events, keeping in mind that under specific conditions, such as low oxygen fugacities and in the presence of metal, carbon and nitrogen might have behaved more like refractory elements than like highly volatile elements, such as noble gases.

In this chapter, we first discuss the origin of the carbon isotopes in the universe and in the solar system. We present estimates of the elemental and isotopic compositions of carbon and, when necessary, of other volatile elements, in various solar system reservoirs. We then discuss the latest estimates of the carbon content and isotopic compositions in the different terrestrial reservoirs. From the comparison between these contrasting inventories, we discuss the various possible processes of delivery of volatiles to the Earth. Finally, the early terrestrial carbon cycle is introduced.

## CARBON IN THE UNIVERSE

### Nucleosynthesis of carbon and stellar evolution

Only hydrogen, helium and some lithium were created in significant amounts in the Big Bang (Table 1). All other elements were formed by nucleosynthesis in stars (e.g., Burbidge et al. 1957; Meyer and Zinner 2006; Truran and Heger 2003), except for most lithium, beryllium and boron that are largely the products of fragmentation (spallation) of heavier nuclei by energetic cosmic rays in the interstellar medium. The build-up of elements (and isotopes) heavier than helium once star formation began in galaxies is often referred to as galactic chemical evolution (GCE). Stars range in mass from the smallest brown dwarfs with mass of ~0.012 solar masses ($M_o$) to super massive stars of >100 $M_o$.

Nucleosynthesis of ever heavier isotopes is the prime energy source for stars and it proceeds in stages that require increasing temperatures and pressures with each new stage. The first stages of nucleosynthesis, deuterium ($D + {}^1H = {}^3He$) and lithium ($^7Li + {}^1H = 2\ {}^4He$) burning, occur even as a star is still forming/contracting (pre-main sequence). Brown Dwarfs, the lowest mass stars (0.012-0.08 $M_o$), do not get beyond the deuterium-burning stage. Higher mass stars join the main sequence, where they spend most of their lives, when they begin burning $^1$H in their cores. Hydrogen-burning converts $^1$H to $^4$He via proton-proton chain and CNO cycle reactions. In the CNO cycle, carbon, nitrogen and oxygen isotopes inherited from earlier generations of stars 'catalyze' the conversion of $^1$H to $^4$He. The isotopes $^{13}$C, $^{14}$N and $^{17}$O are important intermediate products of the CNO cycle.

The main sequence life of a star ends when it has exhausted all the $^1$H in its core. In <10 $M_o$ stars, hydrogen-burning continues in a shell above the core and the star becomes a red giant branch (RGB) star. Massive convection at this transition brings hydrogen-burning products to the stellar surface, most notably $^{13}$C, $^{14}$N and $^{17}$O. Eventually, contraction of the core and mass added to it from the hydrogen-burning shell produce conditions that enable the next stage, core helium-burning. In helium-burning, three $^4$He nuclei are fused together in the triple-alpha reaction, to form $^{12}$C. Further alpha addition leads to the formation of $^{16}$O, but reaction rates beyond this are very slow. During core helium-burning, the entire star contracts and it leaves the RGB. However, once helium is exhausted in the core, shell burning resumes and the star joins the asymptotic giant branch (AGB). For stars of ~3-10 $M_o$, there is a second dredge-up of hydrogen-burnt material into the envelope of the star. The $^{12}$C/$^{13}$C ratios in the envelopes of most stars at the beginning of the AGB are <10, but once alternating phases of hydrogen- and helium-burning in shells above the exhausted core become established, $^{12}$C-rich helium-burnt material is periodically mixed into the envelope. For ≤4 $M_o$ stars, the result of adding this material to the envelop is a rapid increase in both the $^{12}$C/$^{13}$C and the C/O ratios. Eventually, so much $^{12}$C is added that the C/O ratio of the envelope, which was initially less than one, exceeds one. For 4-10 $M_o$ stars, conditions at the base of the envelope allow for some hydrogen-burning, which destroys much of the added $^{12}$C, so that the $^{12}$C/$^{13}$C ratio remains low and the C/O ratio stays below one.

The chemistry of the envelope during the AGB is important because the AGB stars begin to lose their envelopes in massive stellar winds, the development of which is driven at least in part by dust formation. The chemistry of envelopes dictates the types of dust that form – envelopes with C/O<1 are dominated by crystalline and amorphous silicates and oxides, with most of the carbon tied up in CO, while those with C/O>1 are dominated by SiC and graphite and/or amorphous carbon, and most of the oxygen is in CO. These AGB winds are important sources of freshly synthesized material, particularly $^{13}$C and $^{17}$O, and dust for the interstellar medium (ISM).

Eventually, the entire envelope is lost, leaving behind the inert core, now a white dwarf, which is dominated in most cases by $^{12}$C and $^{16}$O. The cores of the most massive AGB stars may experience carbon-burning, which converts $^{12}$C to $^{20}$Ne and $^{24}$Mg, producing an O-Ne-Mg white dwarf. A white dwarf is not always the final state of an intermediate mass star. If it is part of a close binary, it can gain mass from its less evolved companion. The increase in temperature and pressure at the surface as material is accreted eventually results in a nova explosion driven by runaway CNO cycle reactions ($^{12}$C/$^{13}$C<5). However, not all the accreted material is blown off in a nova explosion, and if after repeated nova events the mass of a white dwarf eventually exceeds ~1.4 $M_o$ it will explode as a type Ia supernova.

Once 9-25$M_o$ stars leave the main sequence, their initial evolution is not unlike that of intermediate mass ones, with the development of a supergiant phase and dredge-up that enriches the envelope in CNO-cycle products. However, there is never any dredge-up of He-burnt material, so the C/O<1. Significant mass loss and dust formation is observed to occur during the supergiant phase, but the envelope is never entirely lost before the star becomes a type II supernova. The internal structure of the pre-supernova star is a series of shells that have experienced ever more advanced stages of nucleosynthesis with increasing depth. These include a relatively massive, carbon-rich He-burning shell, but the majority of the shells are oxygen-rich. Two of the major nucleosynthetic products in type II supernovae are $^{12}$C and $^{16}$O. Further nucleosynthesis occurs during the explosion, producing the elements heavier than iron.

For the most massive stars, (>25$M_o$), mass loss during the supergiant phase is so vigorous that the envelope is lost, exposing hydrogen-burnt material directly at the surface. With continued mass loss, the products of hydrogen-burning, such as $^{14}$N, become highly overabundant. The star is now a Wolf-Rayet star. Further mass loss reveals He-burnt material, rich in $^{12}$C and $^{16}$O (C/O>1), and it is only at this stage that dust formation is observed. The ultimate fate of a WR star is a type Ib supernova that produces a similar array of elements/isotopes to type IIs.

**Table 1.** The principal nucleosynthetic formation reactions and sources of the isotopes of hydrogen, carbon and nitrogen.

|        | Nucleosynthesis | Sources        |
|--------|-----------------|----------------|
| $^1$H      | Big Bang        | Big Bang       |
| D      | Big Bang        | Big Bang       |
| $^{12}$C    | He-burning      | SN, AGB        |
| $^{13}$C    | CNO cycle       | AGB            |
| $^{14}$N    | CNO cycle       | AGB            |
| $^{15}$N    | CNO cycle       | Novae, SN, AGB |

SN=supernovae, AGB=asymptotic giant branch stars.

**Galactic chemical evolution**

The lifetimes of stars vary dramatically with their masses. Stars that are only slightly less massive than the Sun have lifetimes that are of the order of the age of the Universe. These low mass stars will not have contributed to GCE and will not be discussed further here. Intermediate and massive stars (>1 $M_o$) have lifetimes that range from ~$10^{10}$ yrs at the low mass end to only a few million years at the high mass end. Early in the history of the Galaxy, massive stars (their winds and type II and Ib supernovae) would have been the main sources of elements/isotopes heavier than helium. The main products would have been isotopes like $^{12}$C and $^{16}$O, so-called primary isotopes because they can be built up from the primary products of the Big Bang (hydrogen and helium). With time and after several generations of massive star formation, the longer-lived intermediate mass stars would have started dying and isotopes like $^{13}$C and $^{17}$O would have begun to build up. Evidence for this GCE can be found in long-lived low mass stars, some of which can have very low abundances of elements heavier than helium (low metallicity). Because rates of star formation decrease with increasing radius from the Galactic center, the effects of GCE can also be seen in the carbon and oxygen isotopic compositions of the interstellar medium, with $^{12}$C/$^{13}$C and $^{16}$O/$^{17}$O ratios increasing more-or-less monotonically with radial distance.

**Carbon in the interstellar medium and the presolar molecular cloud**

Once injected into the ISM, stellar material is subject to a number of energetic processes that modifies it. These include supernova-driven shock waves, cosmic rays and UV irradiation. In the diffuse ISM, this means that complex molecules cannot form and dust is either damaged or destroyed. Most rock-forming elements are condensed in amorphous silicate dust, while carbon is distributed between CO and a poorly characterized carbonaceous dust (Zubko et al. 2004; Draine and Li 2007). In the higher densities of molecular clouds, most of the material is shielded from shock waves and UV irradiation. The interiors of molecular clouds are also much denser and colder than the diffuse ISM – temperatures may reach 10 K or less. As a result, all but the most volatile material is condensed in icy dust grains. Despite these temperatures, a complex chemistry can occur in molecular clouds in the gas phase, on grain surfaces and even within grains (Herbst and van Dishoeck 2009). This chemistry is driven to a large extent by cosmic rays that can penetrate the clouds – activation barriers for reactions between cosmic ray generated ions and other gaseous species are essentially zero, allowing for reactions at even these very low temperatures. Because the reaction temperatures are so low, extreme isotopic fractionations can be generated. Deuterium, for instance, can be enriched by several orders of magnitude in water and simple organic molecules.

**Carbon content and isotopic composition of the solar nebula**

Roughly 99.8% of the total mass of the present solar system is in the Sun. Thus, the general assumption is that the Sun's bulk composition must closely resemble the average composition of the material from which the solar system formed. Except for the brief, early episode that destroyed most of the deuterium and lithium, the outer regions of the Sun are unaffected by nucleosynthesis in the stellar interior (e.g., Geiss and Bochsler 1982). Therefore, spectroscopic measurements of the Sun's photosphere combined with hydrodynamic models of its atmosphere can be used to determine most elemental and even some isotopic abundances (Asplund et al. 2009). To estimate the primordial solar or 'cosmic' composition, small corrections to the present-day photospheric abundances are needed to account for gravitational settling and diffusion of elements that deplete the photosphere over the lifetime of the solar system (Turcotte et al. 1998). Solar photospheric abundances have been measured for most elements, although they continue to be refined (Asplund et al. 2009). In particular, since 2002 there has been a significant revision downwards in the abundances of carbon, nitrogen and oxygen. The composition of the solar wind and Jupiter's atmosphere can also provide useful constraints for the abundances of some elements and isotopes. Tables 2 and 3 list the abundances and isotopic compositions of a restricted number of elements that are particularly relevant here. Fig. 1 compares the isotope variations of hydrogen and nitrogen in the solar system, which show a close affinity between primitive meteorites and the Earth.

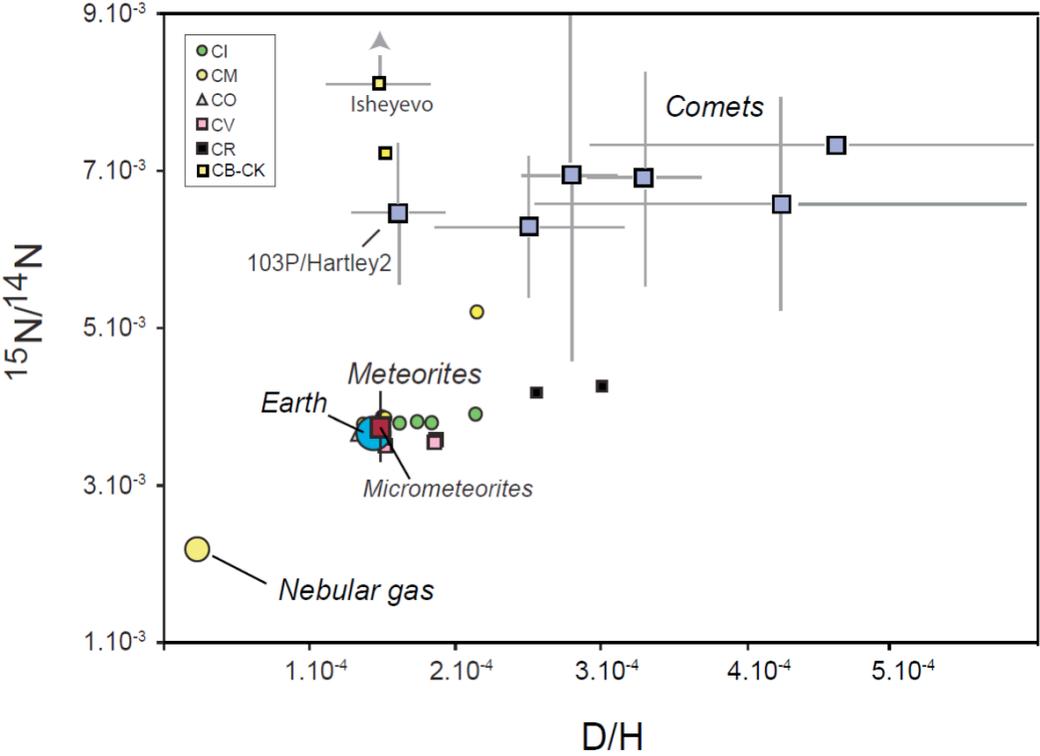

**Fig 1 :** Stable isotope (H, N) variations among solar system objects and reservoirs (adapted from Marty, 2012; data sources therein).

**Table 2.** The abundances (wt.%) of hydrogen, carbon and nitrogen in various reservoirs and objects.

|    | Protosolar[1] | Earth[2] | Halley[3] | CI-CM[4] |
|----|---------------|----------|-----------|----------|
| H  | 71.54 | $3.0 \times 10^{-2}$ | 6.0 | 1.2 |
| He | 27.03 | | | $1.5 \times 10^{-6}$ |
| C  | 0.25 | $5.2 \times 10^{-2}$ | 18 | 3.65 |
| N  | $7.3 \times 10^{-2}$ | $1.7 \times 10^{-4}$ | 0.8 | 0.15 |
| Ne | $1.3 \times 10^{-3}$ | $7.6 \times 10^{-10}$ | | $3.55 \times 10^{-8}$ |
| Ar | $7.8 \times 10^{-5}$ | $5.5 \times 10^{-9}$ | | $1.58 \times 10^{-7}$ |
| Kr | $1.2 \times 10^{-7}$ | $4.6 \times 10^{-10}$ | | $6.25 \times 10^{-9}$ |
| Xe | $1.8 \times 10^{-8}$ | $6.0 \times 10^{-11}$ | | $1.74 \times 10^{-8}$ |

[1] From Asplund et al. (2009). [2] From Marty (2012). [3] From Delsemme (1991) and assuming that the organics has a composition like Halley CHON particles (Kissel and Krueger 1987). [4] From Alexander et al. (2012) and Marty (2012) for noble gases; averages of two carbonaceous chondrites, Murchison (CM) and Orgueil (CI).

**Table 3.** Key isotopic ratios for hydrogen, carbon and nitrogen in various objects.

| | Protosolar | +/- | Earth | Meteorites : chondrites | +/- | Comets : Oort cloud | +/- | Comet : 103P/Hartley2 Kuiper belt | +/- |
|---|---|---|---|---|---|---|---|---|---|
| D/H ($10^{-6}$) | 20[1] | 3 | 149[11] | 120-408[2,6,10] | | 319[4] | 59 | 161[4] | 24 |
| $^{13}C/^{12}C$ ($10^{-2}$) | <1.0[5] | | 1.12 | 1.10-1.14[6,10] | 0.02 | 1.08[9] | 0.10 | 1.05[7] | 0.14 |
| $^{15}N/^{14}N$ ($10^{-3}$) | 2.27[8] | 0.03 | 3.67 | 3.51-5.00[6,10] | 0.2 | 6.80[9] | 1.25 | 6.45[7] | 0.90 |

[1]Geiss and Gloeckler (2003). [2]Robert (2003). [3]Bockelée-Morvan et al. (2008). [4]Hartogh et al. (2011), only $H_2O$. [5]Hashizume et al. (2004). [6]Kerridge (1985). [7]Meech et al. (2011), only CN. [8]Marty et al. (2011). [9]Manfroid et al. (2009), only CN and HCN. [10]Alexander et al. (2012). [11]Lécuyer et al. (1998).

To first order, the solar composition resembles the compositions of young stars in the local neighborhood, as well as the composition of the local interstellar medium (e.g. Nieva and Przybilla 2011). There is scatter in the compositions of other stars in the vicinity of the Sun. In part, this reflects uncertainties in the measurements, but it also reflects spatial and temporal variations in the compositions of the interstellar material from which the stars formed (Kobayashi and Nakasato, 2011). The composition of the interstellar medium is expected to vary in time and space as a result of the infall of material onto the Galactic disk, local 'pollution' by highly evolved mass-losing stars (e.g., supernovae, novae and AGB stars), and because of an overall gradient in star formation rates. Nevertheless, the solar composition is a useful and commonly used reference with which to compare the compositions of other astronomical objects.

Remarkably, the composition of one group of primitive meteorites, the CI chondrites, closely resembles the non-volatile (i.e., not including hydrogen, carbon, nitrogen, oxygen and the noble gases) photospheric abundances. In fact, the resemblance is so close that the appropriately scaled CI abundances of many elements and most isotopes are preferred over the photospheric abundances because they can be measured more accurately (e.g., Asplund et al. 2009; Lodders 2003). The general assumption that the bulk isotopic compositions of the elements in CI chondrites are solar is not valid for the most volatile elements, such as hydrogen, carbon, nitrogen, oxygen and the noble gases. This is because physical and chemical processes in the solar nebula produced several carriers of these elements (e.g., $H_2$, $H_2O$, $CO$, $CH_4$, $NH_3$ and $N_2$, as well as dust/ice) with different volatilities that were physically and isotopically fractionated relative to one another, so that the CI chondrites did not accrete all of these carriers when they formed. Because Solar (or CI) is the starting composition from which all solar system materials are assumed to have ultimately evolved, the compositional variations of solar system objects, particularly meteorites, are often expressed as deviations relative to CI. Here we also use a CI-CM average composition for volatile abundances, since CM-type meteorites such as Murchison are also remarkably primitive and volatile rich, present analogies with micrometeorites (antoher potentially important source of terrestrial volatiles, see below), and might have experienced less aqueous alteration than CIs.

**Volatile abundances and isotope compositions in comets with special reference to carbon**

The volatile species emitted by comets are dominated by water, followed by $CO$, $CO_2$, $NH_3$ and simple organic molecules (Bockelée-Morvan et al. 2004; Mumma and Charnley 2012). To date, there have been no unambiguous remote detections of noble gases in comets. Helium and neon have been measured in cometary matter from comet Wild 2 (or 81P/Wild 2) that was returned to Earth by the Stardust mission (Marty et al. 2008). The isotopic composition of neon more closely resembles that of neon trapped in chondrites (phase "Q" neon) rather than that of solar wind neon, and the helium and neon abundances were found to be very high and only matched by lunar regolith that had been irradiated by the solar wind (SW) for long periods of time. These observations suggest that the light noble gases in Stardust matter were implanted by solar irradiation early in the formation of solar system, provided that phase Q neon is indeed derived from solar neon due to isotope fractionation during implantation as suggested by some models (see Ott 2002, for a review). Hence implantation of solar gases could be a significant source for trace gases, but probably not for carbon that is predominantly in organic compounds.

Based on the data obtained by the VEGA-1 spacecraft as it flew past comet Halley, much of the carbon in that comet is in C-, H-, O-, N-rich (CHON) dust particles (Jessberger et al. 1988; Kissel and Krueger 1987). The C/Mg ratio in Halley dust is 6-12 times that of CI chondrites (Jessberger et al. 1988). This is much higher than for the particles captured by the Stardust mission to comet 81P/Wild 2 (Brownlee et al. 2006). This difference could be largely due to the high relative velocity (6 km/s) between the Stardust spacecraft and the Wild 2 particles at the time they were captured. The particles were captured in a very low-density medium, but the large kinetic energies of the cometary particles would have favoured the preservation of only the more competent and refractory materials. Nevertheless, variations in composition between comets cannot be ruled out.

For comet Halley, Delsemme (1991) estimated a bulk composition of 43 wt.% water, 26 wt.% organics and 31 wt.% silicates, etc. If most of the organics is in CHON-like material (Kissel and Krueger 1987) that is ~70 wt.% carbon, this would mean that Halley-like comets have ~18 wt.% carbon (Table 1). Greenberg (1998) arrived at a similar composition for comets in general using a range of constraints, including the Halley data. This carbon content would correspond to a bulk H/C molar ratio for Halley of about 4, not so different from ratios observed in volatile-rich CI and CM carbonaceous chondrites (molar H/C of 3-13).

**Interplanetary dust particles**

Interplanetary dust particles (IDPs) are small particles that are collected in the upper atmosphere (Bradley 2003). Most IDPs are <50-100 μm across. Larger particles, known as cluster IDPs, tend to be very fragile and fragment on the collectors. Their small size and/or fluffy nature means that IDPs slow down relatively gently during atmospheric entry. Consequently, unlike their larger and/or denser counterparts that at least partially melt (cosmic spherules and micrometeorites) or vaporize, IDPs are less severely heated on entering the atmosphere. Dynamical arguments suggest that IDPs have both cometary and asteroidal sources (e.g., Messenger et al. 2003a).

Most IDPs that have been studied have grossly (factors of 2-3) chondritic elemental compositions, except for volatile elements that can be lost during atmospheric entry heating or are subject to terrestrial contamination (Kehm et al. 2002; Schramm et al. 1989; Thomas et al. 1993). There are extraterrestrial particles with non-chondritic compositions in the collections, but these are more difficult to distinguish from terrestrial contaminants and so are rarely studied. The chondritic particles fall into two morphological groups – porous (CP-) and smooth (CS-) particles. The CP-IDPs are largely anhydrous, composed mostly of submicron crystalline and amorphous silicates, metal, sulfide and organic matter. The silicate compositions tend to be very heterogeneous. Because of their very fine-grained nature, high degree of chemical disequilibrium, and high concentration of organic matter and presolar grains, CP-IDPs are thought to be the most primitive solar system materials available for study. In the CS-IDPs, the silicates are dominated by hydrous minerals (predominantly serpentines and clay minerals). Although not identical, the CS-IDPs resemble the matrices of the aqueously altered chondrites. Because of their resemblance to chondrites and the general belief that comets did not experience aqueous alteration, it is widely assumed that CS-IDPs are derived from asteroids. On the other hand, there are no meteorites that resemble CP-IDPs, which, along with their apparently very primitive nature, has led most researchers to conclude that CP-IDPs are from comets. A cometary origin of at least some CP-IDPs is supported by the high abundances of isotopically anomalous organic matter and presolar grains in them (e.g., Floss et al. 2006; Busemann et al. 2009; Duprat et al. 2010).

On average, the abundance of carbon in IDPs is ~12 wt.% (Thomas et al. 1993), roughly three times that in CI chondrites, although there is considerable variation. Most of this carbon is in organic matter. Messenger et al. (2003b) reviewed the hydrogen, nitrogen and carbon isotopic compositions of IDPs. Their bulk carbon isotopic compositions fall within the terrestrial range, but their average isotopic composition of $\delta^{13}C=-45‰$ is lighter than any bulk meteorites or the bulk Earth (although for the Earth the bulk carbon isotope composition is not precisely known, see below). Hydrogen and nitrogen isotopic compositions vary enormously within and between IDPs. The most striking features of the hydrogen and nitrogen in IDPs are their often very large deuterium and $^{15}N$ enrichments. These enrichments are poorly correlated despite generally being associated with organic matter (See also: Aléon et al. 2003; Keller et al. 2004; Floss et al. 2006; Busemann et al. 2009). The hydrogen and nitrogen isotopic variations are similar to what is seen in the more primitive chondrites (e.g., Floss and Stadermann 2009) and their organic matter (e.g., Busemann et al. 2006) when they are analyzed at similar spatial scales to IDPs.

**Meteorites**

With some rare exceptions, those from Mars and the Moon, meteorites are fragments of asteroids from the asteroid belt located between the orbits of Mars and Jupiter. They are broadly divided into primitive chondrites, and achondrites that have undergone melting and differentiation on their parent bodies (e.g., Scott and Krot 2003). Achondrites generally have low abundances of carbon and other volatiles (hydrogen, nitrogen and noble gases), presumably because their carriers were largely destroyed during melting. The notable exceptions are the ureilites that contain up to ~5-7 wt.% in poorly graphitized carbon and the shock produced high-pressure carbon allotropes diamond and lonsdaleite. The ureilites are a potent potential source of the Earth's carbon. However, their CI-CM-normalized hydrogen and nitrogen are much lower than for carbon. Later on, we will argue that the relative abundances of the Earth's volatiles are CI-CM-like. Hence, if ureilites were a major source of the Earth's carbon, sources that were strongly depleted in carbon relative to the other volatiles must be found. To date, no such sources have been identified. Consequently, the ureilites notwithstanding, here we will concentrate on the chondrites as the most likely sources of the Earth's volatiles.

Historically, the chondrites have been divided into three classes based on their compositions and mineralogies (ordinary, carbonaceous and enstatite). These in turn have been subdivided into a number of groups: ordinary chondrites into H, L and LL, carbonaceous chondrites into CI, CM, CR, CV, CO, CB, CH and CK, and enstatite chondrites into EH and EL. The name carbonaceous chondrite is a historical one and is a bit misleading since some ordinary and enstatite chondrites contain more carbon than some carbonaceous chondrites (Table 4). The chondrite classification scheme is still evolving as more meteorites are found - two new classes (R and K chondrites) have been identified, and a number of individual meteorites do not belong to any recognized group.

After formation, the chondrites experienced secondary modification (thermal metamorphism and aqueous alteration) on their parent bodies. A petrographic classification scheme for secondary processes divides the chondrites into 6 types - types 3 to 6 reflect increasing extent of thermal metamorphism, and types 3 to 1 reflect increasing degrees of aqueous alteration. By convention, the chemical classification is followed by the petrologic one (e.g., CI1, CM2, CV3).

The chondrites are principally made up of three components – chondrules, refractory inclusions and fine-grained matrix – whose relative abundances vary widely (Brearley and Jones 1998). Refractory inclusions and chondrules are high temperature objects, and normally contain little or no carbon. Almost all the carbon in the most primitive chondrites is found in their matrices, as are the other most volatile elements and presolar circumstellar grains. The matrix abundances of these materials in all chondrites are similar to those in CI chondrites (Alexander et al. 2005, 2007; Huss 1990; Huss and Lewis 1994).

Table 4. The range of abundances and isotopic compositions of carbon in bulk and in the directly determined major components – insoluble organic material (IOM) and carbonates (carb) - in the main chondrite groups. Shock heated meteorites were not included in the calculation of the ranges. Tagish Lake is a carbonaceous chondrite that has affinities with the CM and CI chondrites, and fell on January 18$^{th}$, 2000.

|  |  | Min. C (wt.%) | Max. C (wt.%) | Min. $\delta^{13}$C (‰) | Max. $\delta^{13}$C (‰) | Ref. |
|---|---|---|---|---|---|---|
| **CI** |  |  |  |  |  |  |
|  | Bulk | 3.5 | 3.9 | -14.7 | -4.3 | 1 |
|  | IOM | 2 | 2.3 | -17.0 | -17.0 | 2 |
|  | Carb. | 0.014 | 0.420 | 47.8 | 61.0 | 3 |
| **CM** |  |  |  |  |  |  |
|  | Bulk | 1.2 | 3.1 | -19.9 | 3.7 | 1 |
|  | IOM | 0.74 | 1.3 | -34.2 | -8.4 | 2,4 |
|  | Carb. | 0.023 | 0.55 | 3.0 | 67.8 | 3 |
| **CR** |  |  |  |  |  |  |
|  | Bulk | 1 | 2.7 | -11.5 | 0.0 | 1 |
|  | IOM | 0.34 | 1.5 | -26.6 | -20.3 | 2,4 |
|  | Carb. | 0.0023 | 0.048 | 49.9 | 65.4 | 3 |
| **Tagish Lake** |  |  |  |  |  |  |
|  | Bulk | 4 | 4.1 | 9.4 | 14.0 | 1 |
|  | IOM | 1.7 | 1.9 | -14.3 | -13.1 | 5 |
|  | Carb. |  | 1.3 |  | 67.6 | 6 |
| **CV** |  |  |  |  |  |  |
|  | Bulk | 0.27 | 1.1 | -17.1 |  | 1,7 |
|  | IOM | 0.12 | 0.68 | -15.4 | -6.8 | 2,4 |
| **CO** |  |  |  |  |  |  |
|  | Bulk | 0.12 | 0.84 |  | -7.1 | 1,7 |
|  | IOM | 0.04 | 0.48 | -13.9 | -4.5 | 2,4 |
|  | Bulk |  |  |  |  |  |
| **OC** |  |  |  |  |  |  |
|  | Bulk | 0.01 | 0.59 | -22.6 |  | 1,8 |
|  | IOM | 0 | 0.36 | -23.7 | -10.4 | 2,4 |
|  | Carb. | 0 | 1493 | -6.4 |  | 3 |
| **EC** |  |  |  |  |  |  |
|  | Bulk | 0.29 | 0.7 | -14.1 | -4.1 | 9 |

[1]Alexander et al. (2012). [2]Alexander et al. (2007). [3]Grady et al. (1988). [4]Alexander et al. (2010). [5]Alexander et al. (2012). [6]Grady et al. (2002). [7]Kerridge (1985). [8]Grady et al. (1982). [9]Grady et al. (1988).

The main carbonaceous component in chondrites is organic, but aqueously altered meteorites often also contain carbonate (Table 4). The organics in chondrites also accounts for a significant fraction of the hydrogen and most of the nitrogen in the bulk chondrites, as well as being associated with the carrier of most of the noble gases. Carbon is also present in "primitive" meteorites as nanodiamonds that can make up to ~1,000 ppm of the matrix in chondrites (Huss 1990). In bulk, the nanodiamonds contain a trace noble gas component (so-called Xenon-HL) that is clearly of nucleosynthetic origin, so at least some of them are circumstellar. However, nanodiamonds have bulk carbon and nitrogen bulk isotopic compositions that are solar (e.g., Ott et al. 2012), and a dual origin (solar system and inheritance from the presolar molecular cloud) cannot be ruled out. Other trace carbonaceous components include circumstellar SiC and graphite that formed around supernovae, novae and AGB stars prior to the formation of the solar system (e.g., Nittler 2003).

In all chondrite groups, the carbon content decreases with increasing thermal metamorphism, hence the low minimum carbon contents of CV, CO and ordinary chondrites. Generally, little or no carbon remains by the type 3-4 transition, and if any remains much of it may be terrestrial contamination (Alexander et al. 1998). The CK and R chondrites have all been heavily metamorphosed, which is why they have not been included in Table 4. Only in the highly reduced enstatite chondrites is carbonaceous material preserved to higher metamorphic grades (Grady et al. 1986; Alexander et al. 1998). Aqueous alteration can also modify this organic matter, but the effects tend to be subtler (Alexander et al. 2007, 2010; Herd et al. 2011). Hence, to understand the origins of this organic material, here we concentrate on the most primitive chondrites (CI, CR, CM and Tagish Lake) available to us.

**The organic matter in chondrites - relationship to IDPs, comets and ISM**

The organic matter in chondrites can be divided into soluble (SOM) and insoluble (IOM) fractions (Gilmour 2003; Pizzarello et al. 2006). Even in the most primitive meteorites, the concentrations of traditionally defined SOM, that which is soluble in typical solvents (e.g., water, methanol, toluene, etc.), is no more than a few hundred parts per million. The SOM is composed of a very complex suite of compounds that include amino acids, carboxylic acids and polycyclic aromatic hydrocarbons (PAHs). The IOM and carbonates are usually assumed to make up the bulk (>75 %) of the carbon in chondrites. However, the true IOM, that which remains insoluble even after demineralization with acids, only comprises about half of the bulk carbon in primitive chondrites (Table 4). The remainder is in a poorly understood material that is insolvent until hydrolyzed by acids, but it has a composition that is probably not very different from the IOM (e.g., Alexander et al. 2012). For this discussion, it will be assumed that the IOM and the hydrolysable material are closely related.

The most primitive meteoritic IOM has an elemental composition ($C_{100}H_{75}N_4O_{15}S_4$) (Alexander et al. 2007) that is similar to the average composition ($C_{100}H_{80}N_4O_{20}S_2$) of comet Halley CHON particles (Kissel and Krueger 1987). The IOM appears to be composed of small PAHs decorated and cross-linked by short, highly branched aliphatic material (e.g. Cody et al. 2002). At least in terms of hydrogen, nitrogen and carbon isotopes, the bulk compositions and range of compositions seen in the most primitive meteoritic IOM (Busemann et al. 2006) also resembles the IOM in the most primitive IDPs (Aléon et al. 2000; Brownlee et al. 2006; Busemann et al. 2009; Messenger, 2000). The large D and $^{15}$N isotopic enrichments in IOM in chondrites and IDPs are thought to be the result of ion-molecule reactions and other processes in the ISM (Aléon et al. 2003; Messenger 2000; Robert and Epstein 1982; Yang and Epstein 1983) and/or the outer solar system (Aikawa et al. 2002; Gourier et al. 2008). There are variations in the isotopic composition of IOM within and between chondrites and IDPs, but this mostly reflects parent body processing in chondrites (Alexander et al. 2007, 2010; Herd et al. 2011) and atmospheric entry heating of IDPs (Bockelée-Morvan et al. 2004). The organic particles in the Stardust samples also have hydrogen and nitrogen isotopic anomalies (Brownlee et al. 2006; McKeegan et al. 2006; De Gregorio et al. 2010; Matrajt et al. 2008). The hydrogen isotope anomalies in the Stardust samples are more subdued than in the most primitive chondrites and IDPs, but this may reflect modification during capture. A further link between the organics in chondrites, IDPs and comets are the presence of organic nanoglobules in all three (De Gregorio et al. 2010; Flynn et al. 2006; Garvie and Buseck 2006; Matrajt et al. 2012). Thus, the organics in meteorites, CP-IDPs and comets appear to be related despite their parent bodies' very different formation conditions and locations.

The bulk carbon content of CI chondrites, which is largely in organic material, represents ~10% of the carbon available assuming a solar composition for the primordial solar system. The carbon contents of comet Halley and the parent bodies of IDPs may be much higher (30-100% of available carbon), although the uncertainties are large (e.g., Alexander 2005). Nevertheless, whether the organic matter was inherited from the protosolar molecular cloud or formed in the solar system, the formation mechanism was relatively efficient.

## THE SOLAR SYSTEM: DYNAMICS

Despite its apparently well-ordered orbital architecture, the solar system's history is thought to have been dynamic. Indeed, there are at least two phases of evolution during which the Earth could very well have been prevented from forming or have been destroyed. Our understanding of these phases comes from models of orbital dynamics and should thus not be interpreted as absolute truth. Indeed, Alexander et al. (2012) have questioned some of the predictions of the "Nice" and "Grand Tack" models that will be described below. Nonetheless, it is important to realize that simpler models systematically fail to reproduce the observed orbital architecture of the solar system (e.g., Raymond et al. 2009).

During the first dangerous phase, interactions with the gaseous protoplanetary disk may have caused Jupiter to migrate inward to just 1.5 AU, in the immediate vicinity of the Earth's formation zone, before "tacking" and migrating back outward beyond 5 AU (the "Grand Tack" model of Walsh et al. 2011). This excursion of Jupiter into the inner solar system may explain the relatively small mass of Mars. Had the timing of events – particularly of Saturn's orbital migration relative to Jupiter's – been different, Jupiter might well have continued migrating inward and decimated the Earth's building blocks. In the second dangerous phase, the giant planets are thought to have undergone a dynamical instability that re-arranged their orbital architecture and caused the late heavy bombardment (LHB; the instability is referred to as the "Nice" model of Gomes et al. 2005; Tsiganis et al. 2005; and Morbidelli et al. 2007). Although this did cause a large increase in the impact flux throughout the solar system, it was an extremely weak instability: the eccentricity distribution of extra-solar planets suggests that between 50 and 90% of all planetary systems with giant planets undergo much stronger instabilities that would likely have destroyed the Earth, or at least provoked collisions between the terrestrial planets (Raymond et al. 2010, 2011). Even in the current epoch, chaotic dynamics driven by gravitational interactions between planets allow for the possibility of a dynamical instability and future collisions between the terrestrial planets (Laskar & Gastineau, 2009).

We now review the current paradigm of the formation history of the solar system, focusing on the growth of the terrestrial planets and how they acquired their volatiles, notably their carbon. For a more detailed account of the physics of planet formation, the reader is directed to recent reviews on the subject, notably Papaloizou and Terquem (2006) and Morbidelli et al. (2012).

The solar system formed from a disk of gas and dust orbiting the young Sun, sometimes called the "solar nebula" or "protosolar nebula". This disk is assumed to have had the same bulk composition as the Sun, but the composition of the gas and dust phases was not uniform throughout the disk. Driven by both stellar and viscous heating, the disk was hotter closer to the Sun and cooler farther away. Given the temperature-dependent condensation sequence, any given species remains in the gas phase interior to an orbital radius that corresponds to its condensation temperature for the relevant, very low pressure, and should condense beyond that radius. For example, interior to the "snow line," where the temperature is ~170 K, water will remain in the vapour phase, but beyond the snow line it condenses as ice. Likewise, each species in a disk has its own condensation line, exterior to which it will condense and interior to which it remains in the gas phase. The composition of solids that formed within the disk thus reflects the local temperature and the disk's composition; this confluence is called the *condensation sequence* (Pollack et al. 1994; Lodders 2003).

The dominant carbon-bearing condensable species in disks are $CH_4$ and CO (e.g., Dodson-Robinson et al. 2009), although, as noted in the preceding section, carbon-rich dust that is not necessarily seen astronomically may constitute another important, possibly dominant, source of carbon. Polycyclic aromatic hydrocarbons (PAHs) are observed to be abundant in the interstellar medium (Tielens 2008) and in carbonaceous meteorites (see previous section), and may thus constitute an important source of carbon (Kress et al. 2010). Two important orbital radii in terms of carbon abundance have been proposed: a radius analogous to the snow line beyond which carbon can condense – this has been referred to as the "tar line" by Lodders (2004) - and the radius interior to which the PAHs that were initially presented in the disk are destroyed by thermally-driven reactions – this was dubbed the "soot line" by Kress et al. (2010). However, complex, refractory organics, like that seen in meteorites, IDPs and comets, do not form spontaneously via condensation of $CH_4$, CO, etc. at low temperatures. Fischer-Tropsch-type (FTT) synthesis can occur if the appropriate catalysts are present, but there is no evidence that the organics in meteorites, for instance, formed by this mechanism (e.g., Alexander et al. 2007). The IOM does start to break down at temperatures of 300-400°C, thus a more appropriate definition for a "tar line" would be the isotherm beyond which the IOM-like material is stable in the solar nebula.

Protoplanetary disks are not static, but evolve in time. The gaseous component of the disk is dissipated on a timescale of a few million years, as inferred from observations of infrared excesses in the spectra of young stars (Haisch et al. 2001; Meyer et al. 2008). As the disk dissipates it cools, such that the location of the various condensation fronts move inward in time (Sasselov and Lecar 2000; Ciesla and Cuzzi 2006; Dodson-Robinson et al. 2009). The exact positions of the condensation fronts depend on the disk's detailed temperature and pressure structure, which is in turn determined by poorly constrained physical characteristics, such as the viscosity and the opacity (e.g., Garaud and Lin 2007). In addition, dust particles do not remain on static orbits within the disk but migrate radially due to drag forces and pressure, although models suggest that dust migration probably does not produce large pileups (Hughes and Armitage 2010).

Terrestrial planets and giant planet cores form from the dust component of the disk. Giant planets subsequently accrete massive gaseous atmospheres directly from the gaseous component of the disk. Given the strong constraint that gaseous protoplanetary disks only survive for a few million years (Haisch et al. 2001), and the equally strong isotopic constraints that the Earth's accretion lasted for at least 11-30 million years (Yin et al. 2002; Kleine et al. 2002) and possibly more like 50-100 Ma, Touboul et al. 2007), it appears that gas giant planets form faster than the much smaller terrestrial planets. This is counter-intuitive, given that the giant planets' cores are thought to be more massive than Earth (Guillot 2005), and may be explained by the simple increase in the amount of condensable material in the giant planet-forming part of the disk because it lies beyond the snow line (Stevenson and Lunine 1988; Kokubo and Ida, 2002 ). Although the model we present is the current paradigm, we note that there exists an alternate "top-down" model for giant planet formation that invokes gravitational instability in the gas disk (Boss, 1997).

Planet formation occurs in a series of dynamical steps. First, dust grains agglomerate to form mm- to cm-sized pebbles or possibly even m-sized boulders via low-velocity collisions (e.g., Blum and Wurm 2008). Next, planetesimals form from the pebbles/boulders, probably by hydrodynamical processes that efficiently concentrate these particles to a large-enough degree to create gravitationally-bound clumps (Chiang and Youdin 2010; and references therein). The sizes of planetesimals is debated: some formation and collisional models suggest that planetesimals are "born big", with radii of hundreds to a thousand kilometres (Johansen and Youdin 2007; Cuzzi et al. 2008, 2010; Morbidelli et al. 2009; Chambers 2010), while other models argue in favor of much smaller, sub-km sized planetesimals (Weidenschilling, 2011).

The next phase in the growth of solid bodies is the accretion of planetary embryos from planetesimals. A planetesimal that grows larger than its neighbours can rapidly increase its collisional cross section due to gravitational focusing and undergo runaway accretion (Safronov and Zvjagina, 1969; Greenberg et al. 1978; Wetherill & Stewart 1993). Runaway accretion slows to so-called "oligarchic growth" as the orbits of nearby planetesimals are excited by the growing embryo (Ida and Makino 1992; Kokubo and Ida 1998). This process is thought to produce a population of embryos with comparable masses: roughly lunar- to Mars-mass in the terrestrial planet region, and ~Earth-mass in the giant planet region (Kokubo and Ida 2002).

A giant planet core needs to reach at least 5-10 Earth masses before it can efficiently accrete gas from the disk (Ikoma et al. 2000), but the growth of giant planet cores from Mars- to Earth-mass embryos remains poorly understood. In this size range, embryos are far more efficient at scattering planetesimals than accreting them, and standard growth models fail to produce giant planet cores in the lifetime of the gaseous disk (Thommes et al. 2003). However, two proposed mechanisms may help to solve this problem. First, embryos are massive enough to have thin gaseous envelopes, which act to enhance their collisional cross section by a factor of up to ten or more in radius (Inaba and Ikoma 2003). Second, embryos probably experience large-scale orbital migration via both the back-reaction from planetesimal scattering (Fernandez and Ip, 1984; Kirsh et al. 2009) and via tidal interactions with the gaseous protoplanetary disk, often referred to as "type 1" migration (Goldreich & Tremaine 1980). Hydrodynamical models show that there exist locations in the disk where type 1 migration is convergent (Paardekooper et al. 2011; Lyra et al. 2010). A combination of these mechanisms may allow relatively small embryos to rapidly grow into full-fledged giant planet cores (e.g., Levison et al. 2010; Horn et al. 2012), although this remains an area of active research.

Giant planet cores accrete gas from the disk at a rate that is limited by their ability to cool and contract (i.e., their atmospheric opacity), thus freeing up space within their Hill sphere for additional gas (e.g., Hubickyj et al. 2005). Thus, gas accretion initially proceeds at a relatively slow pace; the combination of this time lag and the disk's limited lifetime may explain why giant planet core- (~Neptune)-sized planets amongst exoplanets are so much more common than Jupiter-sized planets (e.g., Howard et al. 2010). Once the mass in a core's gaseous envelope is comparable to the solid core mass, gas accretion enters a runaway phase and the planet becomes a gas giant in just $~10^5$ years (Mizuno 1980). As the planet accretes it clears an annular gap in the disk, which then constrains its accretion rate and final mass (e.g., Lissauer et al. 2009).

Once a giant planet clears a gap in the disk its orbital evolution is inextricably linked to the disk dynamics. This occurs because the gap prevents gas exterior the planet's orbit from interacting viscously with gas interior to the planet's orbit. As the disk viscously evolves and spreads radially, the vast majority of the gas spirals inward onto the star, and a small fraction of the mass spreads to large orbital radii to balance the angular momentum budget (Lynden-Bell and Pringle 1974). In the giant planet forming part of the disk, the gas flow is generally inward such that a gas giant is essentially dragged inward on the disk's viscous timescale of $10^{5-6}$ years in a process called "type 2" migration (Lin & Papaloizou 1986).

Of course, orbital migration of both type 1 and type 2 only occurs in the presence of the massive but short-lived gaseous component of the disk. When the disk dissipates the gas giants have necessarily reached close to their final masses and orbits, but the most dynamic phase of terrestrial planet formation is in full swing.

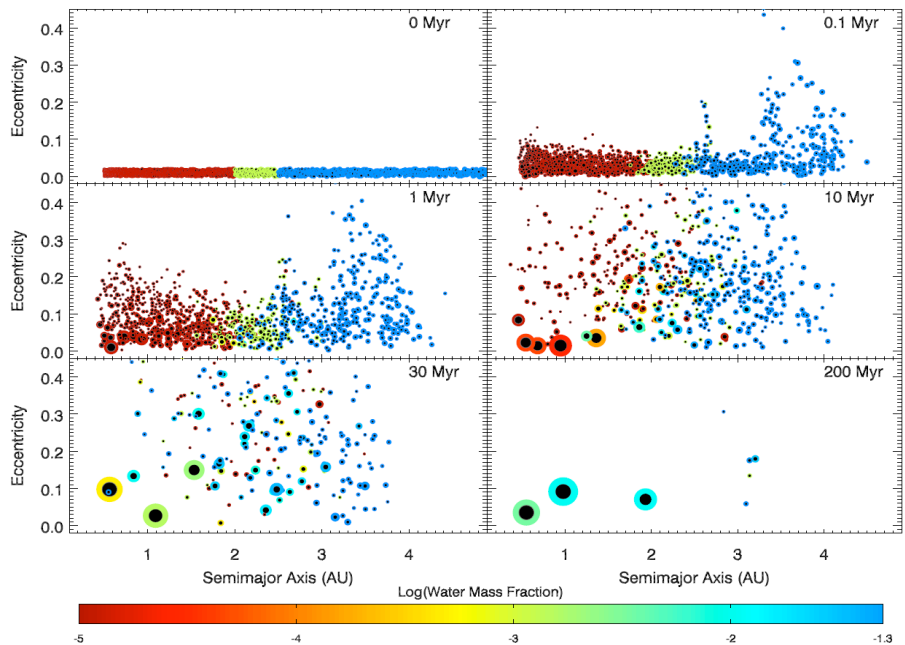

**Fig. 2 :** Six snapshots in time of the orbital configuration of a simulation of late-stage terrestrial accretion from Raymond et al. (2006). The size of each body is proportional to its relative physical size and scales as its mass$^{1/3}$. Jupiter was also included in the simulation on a circular orbit at 5.5 AU (not shown in plot). The color of each particle corresponds to its water content: particles inside 2 AU started the simulation dry, particles that started from 2-2.5 AU have 0.1% water by mass, and particles that started beyond 2.5 AU have 5% water by mass; during collisions the water mass fraction is a calculated using a simple mass balance. In this particular simulation three terrestrial planets formed, as did a few particles trapped in the asteroid belt. The two inner planets are reasonably good Venus and Earth analogues in terms of their masses and orbits, but the outer planet is roughly nine times more massive than the real Mars. The Earth analogue in this simulation accreted a large volume of water of roughly ten times Earth's current water budget, although water loss during impacts was not accounted for. Recall that we expect carbon delivery to Earth to have followed the same dynamical pathway as water delivery.

The final phase in the growth of the terrestrial planets involves the collisional accumulation of planetesimals and planetary embryos. Embryos initially remain isolated from one another as they grow by accreting planetesimals and stay on low-eccentricity orbits via dynamical friction (Kokubo and Ida 1998). Embryos eventually start to interact with each other once the local surface density in planetesimals and embryos is comparable (Kenyon and Bromley 2006). Late-stage terrestrial accretion is thus characterized by slow growth by planetesimal-embryo impacts and punctuated growth by giant embryo-embryo impacts (Wetherill 1985). Each planet's feeding zone spreads outward and widens in time during this chaotic phase of strong gravitational scattering (Raymond et al. 2006; Fig. 2).

A planet's feeding zone determines the mixture of material that condensed in different regions and at different temperatures, and thus the planet's final composition. The temperature at 1 AU is usually thought to have been too hot throughout the disk phase for water to condense (Boss, 1998), although some disk models yield low enough temperatures at 1 AU to allow water to condense (Woolum and Cassen 1999; Sasselov and Lecar 2000). Although it is possible that water vapor could be adsorbed onto grains at 1 AU (Muralidharan et al. 2009), it is generally thought that water must have been delivered to Earth by impacts with icy bodies that formed at cooler temperatures. Primitive outer asteroidal material (C- complex) is a likely source of Earth's water (Morbidelli et al. 2000; Raymond et al. 2004, 2007). C-complex asteroids are thought to be the sources of the carbonaceous chondrites, and of them CI and CM chondrites have compositions that most closely resemble the volatile elemental and isotopic composition of the Earth (Alexander et al. 2012). And it is during an advanced accretion phase that the Earth accreted the bulk of its water (Fig. 2). As we will discuss below, the same arguments hold for Earth's carbon.

Measurements of isotopic chronometers – especially the hafnium-tungsten system – in meteorites and Earth rocks tell us that the Earth's last giant impact occurred at roughly 11-150 Ma (Kleine et al. 2002; Yin et al. 2002; Touboul et al. 2007). This is thought to have been an impact with a ~Mars-sized body, which spun out a disk of material to form the Moon (Benz et al. 1986; Canup and Asphaug 2001). Mars, on the other hand, appears not to have undergone any giant impacts after just a few Ma (Dauphas and Pourmand 2011). Mars may therefore represent a remnant embryo, while Earth is a fully-grown planet that underwent many giant impacts.

After the end of the giant impact phase there remains a population of remnant planetesimals. The final sweep up of these planetesimals occurs by gravitational scattering by the planets, and these planetesimals end up either colliding with the Sun, a terrestrial or giant planet, or getting scattered outward into interstellar space. The fraction of planetesimals that collide with the terrestrial planets provided the so-called "late veneer", which is characterized by the incorporation of highly-siderophile elements (HSEs) into the planets' mantles (Kimura et al. 1974). Given that HSEs are "iron-loving", they should have been sequestered into the core during planetary differentiation, and so their presence in the mantle indicates that they were accreted after the last core formation event or at least time there was significant core-mantle re-equilibration. The late veneer phase represents the tail end of accretion, and the source region of late veneer impactors is constrained by comparing the bulk mantle composition with meteorites (e.g., Dauphas et al. 2004; Burkhardt et al. 2011). At even later times, a very small amount of material was added to Earth during and after the late heavy bombardment (Gomes et al. 2005, Bottke et al. 2012).

Fig. 2 shows the evolution of a dynamical simulation of late stage terrestrial accretion from Raymond et al. (2006). In the simulation, accretion transitions from the primordial accretion to the late veneer after the last giant impact energetic enough to stimulate core-mantle equilibration (and thus reset the hafnium-tungsten chronometer). In this simulation the last giant impact occurred at ~60 Ma for the Earth analog planet (and 20 Ma for the Venus analog).

Accretion simulations like the one from Fig. 2 have succeeded in reproducing a number of characteristics of the actual terrestrial planets, such as their number and approximate total mass (Wetherill 1990) and their low orbital eccentricities and inclinations (Raymond et al. 2006, 2009; O'Brien et al. 2006; Morishima et al. 2008, 2010). In addition, these simulations were able to explain the origin and isotopic signature of Earth's water (Morbidelli et al. 2000; Raymond et al. 2007).

However, these simulations failed to produce realistic Mars analogs. Planets at Mars' orbital distance were systematically a few to ten times more massive than the real Mars (0.11 Earth masses; Wetherill, 1991; Raymond et al. 2006, 2009; O'Brien et al. 2006). Improvements in numerical resolution and the inclusion of additional physical effects have been unable to solve the Mars problem (Morishima et al. 2010).

The only successful solution to the Mars problem came from a change in the "initial conditions". The aforementioned simulations started from disks of planetesimals and embryos that stretched from a few tenths of an AU out to 4-5 AU, just inside Jupiter's current orbit. Mars formed big simply because there was too much mass in the Mars' feeding zone and no mechanism to clear out that mass. However, Hansen (2009) showed that if the terrestrial planets only formed from a narrow annulus of embryos from 0.7-1 AU, a small Mars is produced naturally as an edge effect. Earth and Venus are big because they formed within the annulus of embryos, and Mars and Mercury are small because they were scattered beyond the annulus, thus limiting their growth. In fact, Hansen (2009) was able to reproduce the masses and orbital configurations of all of the terrestrial planets. The only problem was that his initial conditions were completely ad-hoc.

Walsh et al. (2011) proposed a new model called the "Grand Tack", which provides a solution to the Mars problem while remaining consistent with the large-scale evolution of the solar system (Fig. 3). The model relies on the fact that, because the giant planets form much faster than the terrestrial planets, they can sculpt the disk of terrestrial embryos. Figure 4 illustrates the Grand Tack model.

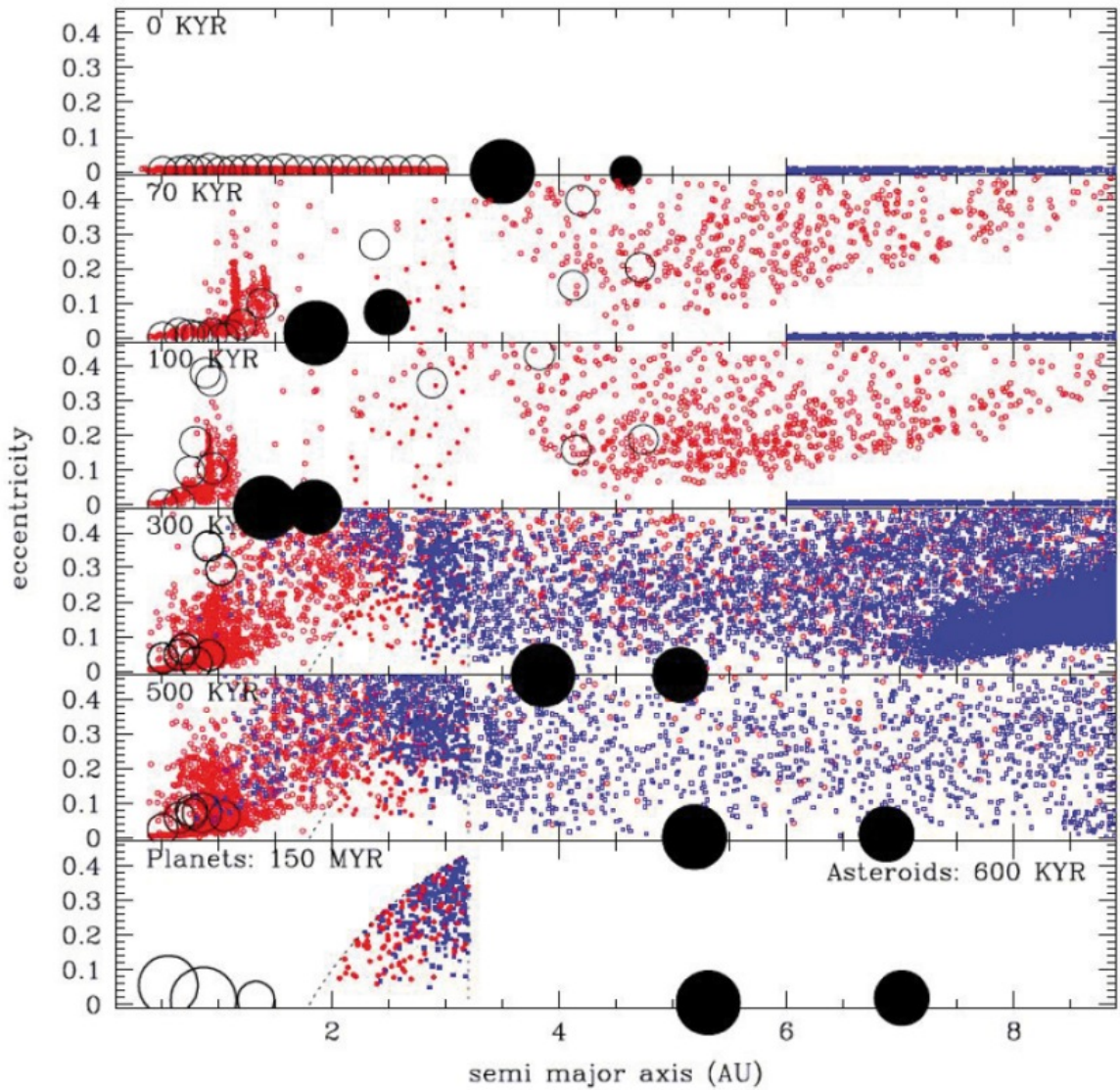

**Fig. 3 :** Snapshots in time of a simulation that represents the Grand Tack model (Walsh et al. 2011). The simulation starts with three parts: (i) an inner disk of planetary embryos (open circles) and dry (red) planetesimals, (ii) a fully-formed Jupiter at 3.5 AU and Saturn's core trapped in 2:3 resonance (large black dots), and (iii) a pristine outer disk of water-rich planetesimals. Jupiter and Saturn migrate inward together while Saturn continues to grow by accreting gas from the disk. Once Saturn reaches its final mass the two planets' migration reverses direction and they migrate outward until the disk dissipates and leaves them stranded at 5.4 and 7.1 AU. (See Walsh et al. 2011 for a discussion of plausible orbital histories of Jupiter and Saturn). The inner disk is truncated at 1 AU and the terrestrial planets that formed are similar to the actual ones, including Mars. Water and carbon are delivered to the Earth in the form of outer disk planetesimals that "overshoot" the asteroid belt as the are scattered inward by Jupiter during the giant planets' outward migration. They can be seem in the 300 and 500 kyr panels as the blue symbols at low semimajor axis and generally high eccentricity.

The Grand Tack model starts late in the gas disk phase, when Jupiter was fully formed but Saturn had not yet grown to its final size. Jupiter migrated via the type 2 mechanism inward from its formation zone, assumed to be at a few AU, perhaps just beyond the snow line. Meanwhile, Saturn started to accrete gas and migrated inward rapidly, catching up with Jupiter. Hydrodynamical simulations show that Saturn is naturally trapped in the 2:3 mean motion resonance with Jupiter (Pierens and Nelson 2008). When the two planets are locked in this configuration, their direction of migration is switched to outward (Masset and Snellgrove 2001; Morbidelli and Crida, 2007; Pierens and Nelson 2008; Pierens and Raymond, 2011). Jupiter and Saturn migrate outward until the disk dissipates, stranding them close to their current orbits.

If Jupiter's "tack" – i.e., change in the direction of its migration – occurred at ~1.5 AU, then the inner disk of embryos is naturally truncated at about 1 AU. The terrestrial planets that form from this disk are a quantitative match to the real ones (Walsh et al. 2011). Although Jupiter migrates through the asteroid belt twice (once going inward, once outward), it is re-populated by two distinct sources. The inner belt contains planetesimals that originated interior to Jupiter's orbit and were subsequently scattered outward during Jupiter's inward migration, and then back inward during Jupiter's outward migration. The efficiency with which inner disk planetesimals are implanted into the final asteroid belt is only ~0.1%, but this is actually in very good agreement with the current belt. The outer asteroid belt contains planetesimals that originated beyond Jupiter's orbit – either between the giant planets or beyond the ice giants' orbits – and were scattered inward during outward migration and implanted with an efficiency of ~1%. Thus, in the context of the Grand Tack model, S-complex asteroids represent leftovers from the terrestrial planet region and C-complex asteroids are leftovers from the giant planet region.

Water delivery to Earth occurs in the Grand Tack model in the form of C-complex material that is scattered inward past the asteroid belt during Jupiter's outward migration (Walsh et al. 2011). For every C-complex planetesimal that was implanted into the main belt, roughly 10-20 planetesimals were scattered onto orbits with roughly the same semi-major axis but with eccentricities high enough to intercept the inner annulus of embryos. This C-complex material therefore represents a "pollution" of the inner annulus of embryos by C-complex planetesimals at the few percent level. Accretion simulations including this tail of C-complex material show that it does indeed deliver about the right amount of water to Earth. In the context of the Grand Tack model, carbon was delivered to Earth by C-complex material, along with the water and other volatiles (see next section).

To conclude this section, we emphasize that although studies have mainly focused on the delivery of water to Earth, the delivery of carbon represents a problem of comparable magnitude. Carbon chemistry in protoplanetary disks is complex, and the dominant species may have been $CH_4$, CO, refractory carbonaceous dust, and/or perhaps free PAHs. However, since carbonaceous chondrite meteorites are thought to represent C-complex asteroids, and since these represent the probable source of water on Earth, carbon probably followed the same path as water in its delivery to Earth.

## CLUES TO THE ORIGIN OF CARBON ON EARTH

### Terrestrial carbon inventory

Making an inventory of terrestrial carbon is not an easy task. The sizes of the main surface reservoirs of carbon (carbonates and reduced carbon in the biosphere and in hydrocarbons), are well documented. In contrast, the inventory of the deep carbon reservoirs is not known, so that estimates require calibration relative to other geochemical proxies and are somewhat model-dependent. Estimates of the carbon bulk Earth content vary by over an order of magnitude (e.g., 50 ppm carbon for Zhang and Zindler 1993; ≥500 ppm carbon for Marty 2012). These are nevertheless low compared to those of potential cosmochemical contributors (comets: 180,000 ppm; carbonaceous chondrites: 36,000 ppm, Table 1). Below we review the carbon content of the major terrestrial reservoirs.

*Earth's surface*

The major reservoirs of carbon at the surface of our planet are carbonates in marine and continental sedimentary rocks, and reduced carbon in the biosphere and in fossil hydrocarbons. Carbonates occur as continental massive units from ancient oceanic platforms, as oceanic sediments, and as veins and alteration phases in the oceanic crust. The amount of reduced carbon is lower than the carbonate reservoir by about a factor of ~4-5. Current estimates of the total amount of carbon at the Earth's surface are around 7-11 x $10^{21}$ moles (Hayes and Waldbauer 2006; Holser et al. 1988; Javoy et al. 1982; Ronov and Yaroshevskhiy, 1976). Carbonates have $\delta^{13}C$ values of around 0 ‰ relative to PDB (the standard PDB is a cretaceous carbonate) and terrestrial organic carbon is more depleted in $^{13}C$, with values around -30 ‰, which is a signature of its biogenic origin. Consequently, the bulk $\delta^{13}C$ of the surface inventory is around -5 to -10 ‰. The composition of this surface carbon may not have varied greatly in the geological past, since ancient sedimentary rocks have carbonate and organic carbon values in the same range (Hayes and Waldbauer 2006). This bulk value is also surprisingly comparable to estimates of the bulk mantle value of -5 ‰: the surface inventory is a mixture of two pools of carbon with different isotopic compositions, one of which being of presumably biogenic origin. This similarity and the constancy of the bulk surface value with time seems to indicate that the relative proportions of carbonates and reduced carbon have been about the same throughout the geological history, which has important implications both for the exchange of carbon between the surface and the deep Earth, and for the size of the biosphere through time.

*Mantle*

The carbon content of mantle-derived material is generally very low because carbon is extensively outgassed from lavas. Thus, current estimates of the mantle carbon content are indirect and based on the calibration of carbon to specific geochemical tracers whose geochemical cycles are well constrained. Here we review current estimates of the carbon content of the mantle. All of them are based on the assumption that carbon is incompatible during partial melting of the mantle source, that is, it goes quantitatively to the surface during magma generation and volcanism.

The highest rate of magma production takes place along divergent plate margins at mid-ocean ridges, which are also the locus of most of mantle degassing. Thus, most estimates are for the mantle source of such magmas, referred in the literature as the depleted mantle (DM). The rare isotope of helium, $^3$He, is of primordial origin in the Earth, that is it was trapped within the forming Earth from a cosmochemical reservoir and it is still degassing from the mantle at present. Its abundance in the atmosphere is low as helium escapes to space. The flux of $^3$He from the mantle has been quantified from excesses of $^3$He in deep-sea waters (knowing the residence time of the latter). The original estimate of 1000±300 moles $^3$He/yr (Craig et al. 1975) has been recently revised downward to 527±106 moles/yr (Bianchi et al. 2010). The C/$^3$He molar ratio measured in mid-ocean ridge basalts (Marty and Jambon, 1987; Marty and Tolstikhin 1998) and in mid ocean ridge hydrothermal vents is on average 2.2±0.6 x 10$^9$. Together with an average partial melting rate of 12±4 % for mid ocean ridge basalts (MORBs) and a total magma production rate of 21 km$^3$/yr at all mid-ocean ridges, one obtains a carbon content of the MORB mantle source of 27±11 ppm carbon. Another approach is to analyse samples from the mantle that are not degassed, or for which degassing fractionation can be corrected for, and to calibrate carbon relative to a non-volatile trace element, such as niobium. Saal et al. (2002) and Cartigny et al. (2008) estimated the $CO_2$/Nb ratio of the DM at 240 and 530, respectively, resulting in estimated carbon contents for the DM of 19 ppm and 44 ppm, respectively. Hirschmann and Dasgupta (2009) and Salters and Stracke (2004) estimated the carbon content of the DM at 14±3 ppm and 16±9 ppm, respectively, using similar approaches. All these estimates suggest a carbon content for the DM of ~20-30 ppm, close to the bulk Earth estimate of Zhang and Zindler (1993) of 50 ppm.

Noble gases released at centers of hot spot volcanism (e.g., Hawaii, Yellostone, Réunion) that are fed by mantle plumes have isotopic compositions suggesting that they come from a less degassed region of the mantle. Thus, the DM is unlikely to represent the bulk Earth inventory. Potassium-40 is produced by the radioactive decay of $^{40}$K, with a half-life of 1.25 Ga, and the total amount of radiogenic $^{40}$Ar produced over 4.5 Ga from terrestrial potassium can be readily computed. A known fraction of $^{40}$Ar is now in the atmosphere and the complementary amount of radiogenic $^{40}$Ar trapped in the silicate Earth is obtained by mass balance. For a bulk silicate

Earth (BSE) potassium content of 280±120 ppm (2 σ) (Arevalo et al. 2009), about half of the $^{40}$Ar produced over the Earth's history is in the atmosphere, and the other half is therefore still trapped in the silicate Earth (Ozima and Podosek 2002; Allègre et al. 1996). The carbon content of the BSE can be scaled to $^{40}$Ar using the composition of gases and rocks from mantle plume provinces following the method presented in Marty (2012), which gives 765 ppm carbon for the BSE (with a large uncertainty of 420 ppm at the 2 σ level when all errors are propagated).

An independent estimate of the BSE carbon content can be obtained using primordial noble gases. Extinct and extant radioactivities producing noble gas isotopes indicate that most of the non-radiogenic noble gases, like $^{36}$Ar, are in the atmosphere, and that their mantle abundance are order(s) of magnitude lower. The mean C/$^{36}$Ar ratio of carbonaceous chondrites is 5.6±2.0 x 10$^7$ (C and $^{36}$Ar data from Otting and Zähringer 1967) and is not significantly different from the ratio for the much less volatile-rich enstatite chondrites (5.9±4.2 x 10$^7$, id.). Assuming a chondritic source and from the $^{36}$Ar atmospheric inventory, one can compute a bulk Earth carbon content that is 920±330 ppm, which is within the uncertainties of the carbon content computed above using a different approach.

These independent estimates suggest that the terrestrial carbon content of the Earth is within 500-1000 ppm. Such a content would correspond to the contribution of about 2±1 wt.% carbonaceous chondrite (CI-CM) material to a dry proto-Earth (Fig. 4). If this estimate is correct, there are several important implications for the origin of volatiles on Earth.

(i) There exists large reservoir(s) of carbon in the Earth that are not yet documented. The occurrence of gas-rich regions of the mantle is supported by the isotopic compositions of mantle plume-derived noble gases (helium, neon) isotope, which point to a mantle reservoir that is less degassed that the DM one;

(ii) Most (>90 %) of the carbon is not at the Earth's surface, but is trapped the bulk silicate Earth, contrary to the case of noble gases that are mostly in the atmosphere;

(iii) The contribution of ~2 wt.% CI material for carbon and other volatile elements (Fig. 4) is higher than the so-called "Late Veener" of about 0.3 wt.% (range : 0.1-0.8 wt.%) of chondritic material to account for the mantle inventory of platinum group elements that never equilibrated with metal, presumably because they were added after the main episodes of core formation (Kimura et al. 1974).

*Carbon in the core ?*

From experimental data, it has been proposed that carbon may be present in the core at the percent level (Dasgupta and Walker 2008; Wood 1993). A high carbon content of the core would change dramatically the carbon inventory of the Earth, and also the bulk carbon isotope composition of our planet since carbon stored in the core would be isotopically different/fractionated with respect to mantle carbon. For instance, a carbon content of 2 % (Wood 1993) in the core corresponds to a BSE content of 6600 ppm carbon. If the source of this carbon was chondritic, then it should have been accompanied by ~ 6 x $10^{16}$ moles of $^{36}$Ar (using the chondritic C/$^{36}$Ar from the preceding subsection), which is an order of magnitude higher than the current estimate of the terrestrial inventory. One could argue that the noble gases were largely lost to space from the primitive atmosphere, and that only the reactive volatiles, such as carbon and, presumably, nitrogen and hydrogen, were quantitatively trapped in the Earth. However, the same argument against extensive sequestration of carbon into the core also applies to the halogens (Sharp et al. 2009). The chlorine content of the bulk Earth is 1.1 x $10^{23}$ g, or 3.6 x $10^{-7}$ mol $^{35}$Cl/g. Because Cl is not siderophile (Sharp et al. 2009), chlorine is in the mantle, in the crust, and in the oceans. The average carbonaceous chondrite $^{35}$Cl/$^{12}$C ratio is ~200. The equivalent carbon concentration of a chondritic Earth is 800 ppm, in fact remarkably comparable to the carbon content computed in the preceding subsection from the $^{40}$Ar budget and from the C/$^{36}$Ar ratio. The isotopic composition of terrestrial chlorine is similar to that of chondrites, and there is, therefore, no reason to suspect that chlorine has been lost to space in early times. Whether or not the amount of carbon in the core is low is an open question that requires further metal-silicate partitioning experimental studies involving not only carbon, but also nitrogen, hydrogen and noble gases.

**Volatile (C-H-N-noble gas) elemental and isotopic constraints**

Carbon isotopes are the primary source of information on the origin of this element in Earth, but the message is ambiguous since carbon has only two isotopes and, because it is a light element, the extent of its fractionation can be high and can superimpose isotope fractionation effects to source heterogeneities. The stable isotope compositions of other volatile elements, such as hydrogen (water) and nitrogen and the associated noble gases, are important supplementary sources of information on the origin of terrestrial carbon, and are discussed in this section.

The range of $\delta^{13}C$ values of mantle carbon, based on the analysis of diamonds, MORBs, mantle plume rocks and volcanic gases is thought to be around -5±3 ‰ (Deines 1980; Des Marais and Moore 1984; Javoy et al. 1986; Marty and Zimmermann 1999). Lighter (more negative) values have been reported for a few MORB glasses and volcanic gases and have been interpreted as being due to isotope fractionation during degassing (Javoy et al. 1986). Lighter isotopic compositions in some of diamonds have generally been interpreted as due to isotopic fractionation or incorporation of recycled organic carbon (Cartigny et al. 1998, 2001; Deines, 1980). Because a large amount of carbon may be sequestrated in the core, a firm carbon isotope budget of the Earth is not yet possible. However, the isotope fractionation of carbon between metal and silicate at the very high temperatures and pressures that prevailed during core formation are unlikely to have been larger than a few per mil.

Terrestrial carbon might have been supplied by the following cosmochemical reservoirs (see also section above).

(i) The protosolar nebula. Solar-like neon is present in the mantle and indicates that a solar-like source supplied a fraction of the terrestrial volatile elements. Whether such a source also supplied carbon to the Earth is an open question. From the analysis of solar wind implanted in lunar regolith grains, Hashizume et al. (2004) proposed that the solar wind isotopic composition is light, with an upper limit for $\delta^{13}C$ of -105±20 ‰. The solar wind may be enhanced in the light isotopes due to Coulomb drag, so that the solar value could be heavier than this limit by about 20-30 ‰ per mass unit for light elements like carbon, nitrogen or oxygen (Bodmer and Bochsler 1998). Thus the solar carbon isotope composition could still be light, possibly lighter than - 80 ‰. If this is correct, then the presence of solar carbon at depth should be reflected by a light carbon component for samples of deep mantle origin, like those associated with mantle plumes. Such an isotopically light carbon component has yet to be observed: mantle carbon shows values of around - 5 ‰ whatever the source is, either MORB-like or plume-like. D/H and $^{15}N/^{14}N$ isotope systematics could only allow for a small (<10 %) fraction of terrestrial hydrogen to be of solar origin (Marty 2012; Alexander et al. 2012).

(ii) Comets. The D/H ratios of these bodies are generally 2-3 times the terrestrial value (Bockelée-Morvan et al. 1998; 2008, Mumma and Charnley 2012), which led to the consensus view that ocean water, and, by extension other major volatiles, like carbon and nitrogen, cannot be derived solely from comets. Recently, a Jupiter family comet, presumably originating from the Kuiper belt, has been shown to have a terrestrial-like water D/H (Hartogh et al. 2011), although Alexander et al. (2012) argue that the bulk D/H of this comet may be significantly higher if it contains Halley-like abundances of organic matter and this organic matter has a D/H ratio like that in the most primitive meteorite organics. $^{15}N/^{14}N$ ratios of comets that have been measured to date are all much higher ($\delta^{15}N > 800$ ‰, where the referecne is atmospheric nitrogen; Arpigny et al. 2003; Bockelée-Morvan et al. 2008; Mumma and Charnley 2012) than the range of terrestrial values ($\delta^{15}N$ from -30 to +40 ‰ for the most extreme end-members), which would preclude a major cometary contribution for terrestrial nitrogen. However, there may be a caveat: the nitrogen isotopic compositions have only been measured in cometary CN and HCN, and the composition of cometary $NH_3$, $N_2$ (if quantitatively trapped) and refractory organic matter are not known.

(iii) Chondrites. Their $\delta^{13}C$ compositions are generally negative, with values of around 0 to -15 ‰ (Table 4), that encompass the terrestrial carbon isotopic composition and, therefore, support a chondritic origin for this element. The hydrogen and nitrogen isotope ratios of the Earth are best matched by the bulk CI chondrites, followed by the CMs, with a small input of solar material (Marty 2012; Alexander et al. 2012).

There is one notable exception for a chondritic source for terrestrial volatiles, and that is the isotopic composition of atmospheric xenon. Atmospheric xenon isotopes and, to a lesser extent, krypton isotopes are enriched by 3-4 % per amu (about 1 % per amu for krypton), relative to chondritic or solar compositions. Xenon is also depleted in its elemental abundance by a factor of about 20 relative to other noble gases in the atmosphere (compared to a chondritic abundance pattern). This fact, known as the xenon paradox, has been regarded as the result of early escape processes specific to the physical properties of xenon (Dauphas 2003; Pepin 1991; Tolstikhin and O'Nions 1994). Recently, Pujol et al. (2011) reported xenon isotopic compositions in Archean rocks that are intermediate between atmospheric and chondritic ones, and suggested a long term, non-thermal escape of atmospheric xenon through time due to ionization by solar UV light. It is not known whether such a process could have affected carbon at the Earth's surface. Its effect might have been limited if carbon is indeed mostly sequestrated in the mantle. The possibility of xenon loss from the atmosphere after the Earth's forming events has implications for the timing of early atmospheric evolution and of Earth's formation (see below).

**Inferences on the nature of Earth's building blocks**

According to our estimate, the delivery of volatile elements, including carbon, to the Earth required the contribution of 1-3 wt.% of "wet" material, having probable affinities with CI, or possibly CM, chondrites. By comparison, platinum group elements, which are in chondritic proportions in the mantle, require a lower contribution of about 0.3 wt.% (range: 0.1-0.8 wt.%) of chondritic material after core formation. This difference suggests that the delivery of volatile elements was not a late veener event and was already on-going during terrestrial differentiation. The depletion of nitrogen in the BSE (Fig. 4), if due to nitrogen trapping in the core, also requires early accretion of volatiles. This point is important for understanding the nature of the building blocks of our planet. The ruthenium and molybdenum isotope signatures of most meteorite groups correlate (Dauphas et al. 2004; Burkhardt et al, 2011) and show enrichments in s-process isotopes relative to Earth's. Molybdenum is moderately siderophile, while ruthenium is highly siderophile. Thus, mantle molybdenum was predominantly accreted during the main stage of planetary growth, as were volatile elements, while the ruthenium now present in the mantle was delivered as part of a late veneer. Thus the respective cosmochemical reservoirs of these deliveries must have had comparable isotopic compositions during and after Earth's accretion. While most meteorite groups show enrichments in s-process molybdenum and ruthenium isotopes relative to Earth, the exceptions are the enstatite and CI (Orgueil) chondrites for molybdenum and, presumably, for ruthenium, which both have terrestrial-like molybdenum isotopic patterns. Hence making the Earth from a cosmochemical reservoir having an enstatite-like isotope composition (which has the appropriate oxygen isotope signature) with addition of "wet" material akin to CI-like material is not only compatible with volatile element compositions, but also could have provided a way to oxidize the composition of the Earth. Importantly, if the volatile elements were accreted before the end of terrestrial differentiation as we argue here, a major fraction of them must have survived the Moon forming impact and other giant impacts that built the early Earth. An open question of this scenario is the behaviour of carbon and other volatile elements during core formation if they were delivered not as a late veneer but during terrestrial differentiation. Much progress is expected from experimental HP-HT works on volatile elements in the presence of metal and silicate.

**Cosmic dust as a major source of terrestrial volatiles ?**

Micrometeorites (MMs) and interplanetary dust particles (IDPs) are potentially important suppliers of terrestrial volatiles because (i) the D/H distribution of MMs' values presents a frequency peak that coincides with that of terrestrial water (Engrand et al., 1999), (ii) at present, their mass flux of 10,000-40,000 tons/yr (Love and Brownlee, 1993; Engrand and Maurette, 1998) exceeds by far that of meteorites (~8 tons/yr; Bland et al., 1996), and (iii) "cosmic" dust may be rich in carbon (e.g., Matrajt et el., 2003). Marty et al. (2005) estimated that <10 % atmospheric nitrogen could have been supplied by cosmic dust after the Earth's building events since 4.4 Ga ago, with essentially the correct nitrogen isotopic composition, based on the analysis of nitrogen in MMs and using the lunar surface record of ET contribution to planetary surfaces. The impact on the carbon inventory of our planet might have been even more limited since nitrogen is apparently depleted by one order of magnitude in earth relative to carbon (Fig. 4).

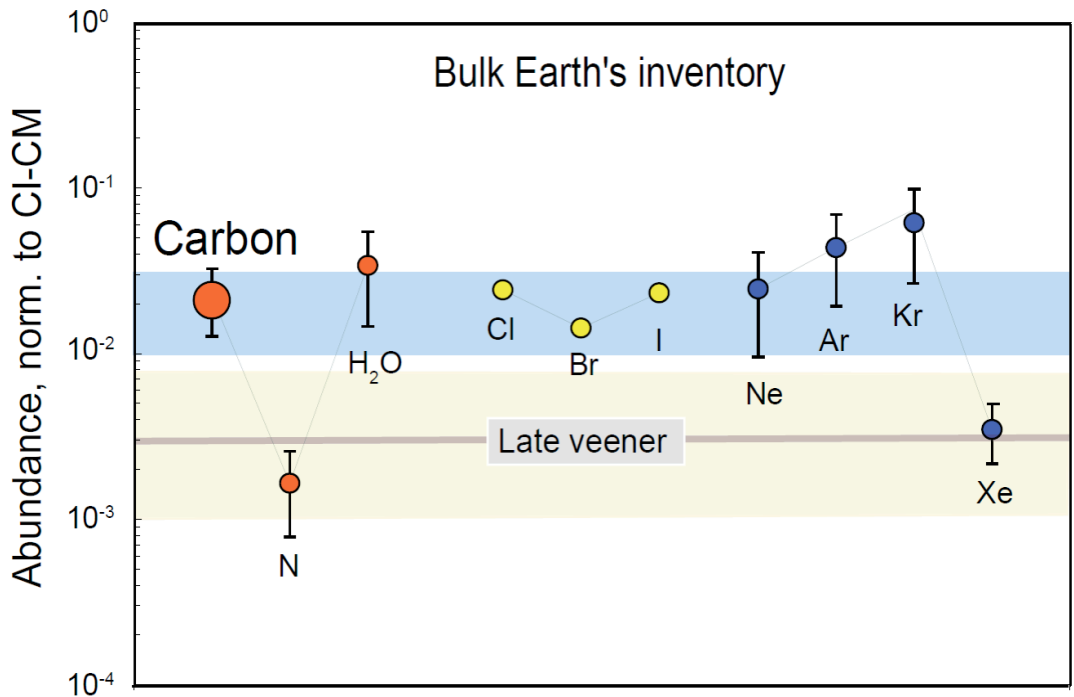

Fig. 4 : Normalised abundance of volatile elements in the Bulk Earth (data from Marty, 2012). Each volatile molar abundance is divided by the mass of the Earth (5.97 x $10^{27}$ g) and normalized by the corresponding content in CI-CM chondrites. Note that the terrestrial abundance pattern is close to chondritic and corresponds to the contribution of ~2±1 % carbonaceous chondrite material. Such contribution is in excess of the "Late Veener" one, corresponding to the amount of highly siderophile elements in the mantle.

From a dynamical point of view, IDPs do not represent a plausible source for the bulk of Earth's Carbon. During planetary growth, dust is efficiently accreted by planetesimals and planetary embryos (Leinhardt et al 2009, Lambrechts & Johansen 2012) such that its lifetime during the main phases of planetary accretion is short. Dust is re-generated by collisions, but is swept up by larger bodies on a timescale that is short compared with the radial drift timescale (Leinhardt et al 2009). It is only after the "dust has settled" in terms of planet formation, and the density of planetesimals and surviving embryos (a.k.a. terrestrial planets) has decreased sufficiently in the inner Solar System, that dust particles are able to drift over large radial distances and thus deliver volatiles such as Carbon and water to Earth. These arguments would break down if the snow-/soot-line swept in to ~1 AU during the last stretch of the gaseous protoplanetary disk phase (e.g., Sasselov & Lecar 2000), producing volatile-rich dust grains close to Earth's orbit. However, it has not been demonstrated that such a scenario could deliver the requisite amount of volatiles to Earth.

### *Timing of volatile delivery and retention*

The time constraints for this epoch are scarce and somewhat model-dependent. Inner planet bodies of the size of Mars accreted in a few Ma (Dauphas and Pourmand 2011). It takes longer to make an Earth-size planet, but how long is a matter of debate. The $^{182}$Hf-$^{182}$W extinct radiochronometer permits one to explore metal-silicate differentiation of planetary bodies, given the half-life of $^{182}$Hf (9 Ma) and the lithophile/siderophile nature of Hf/W. The Earth has a different W isotope composition compared to chondrites, which allows one to set the last episode of metal-silicate differentiation, presumably the formation of the core, at 11-30 Ma after start of solar system condensation (e.g., Yin et al. 2002; Kleine et al. 2002). The final stage of core formation has been attributed by many to the Moon forming impact, after which no similarly catastrophic collisions would have taken place. Touboul et al. (2007) have proposed a longer timeframe (60-120 Ma) since they could not find any W isotopic difference between the Earth and Moon. They argued that, because the terrestrial and lunar mantles have different Hf/W ratios, the absence of any W isotopic difference between the two bodies implies that their last major episode of differentiation (the Moon forming impact) took place after $^{182}$Hf was decayed, in practice, ≥50 Ma. However, this assumption would not hold if the Earth and Moon's mantles have a similar Hf/W ratio, which could have been homogenized between the terrestrial and lunar mantles during or in the aftermaths of the Moon forming impact.

Independent time constraints for Earth formation come from the notion of atmospheric closure using the extinct radioactivity of $^{129}$I (this isotope decays to $^{129}$Xe with a half life of 16 Ma). The initial abundance of $^{129}$I can be estimated from that of terrestrial iodine (the stable $^{127}$I isotope) and from the amount of $^{129}$Xe in the atmosphere in excess of the stable Xe isotope composition. This mass balance yields a closure age of about 120 Ma after start of solar system formation for the atmosphere (Wetherill, 1975), which is also that of the bulk Earth since the mantle inventory of radiogenic $^{129}$Xe is small compared to the atmospheric one (e.g., Ozima and Podosek, 2002). Combining this radiochronometric system with the other relevant extinct radioactivity of $^{244}$Pu ($T_{1/2}$ = 82 Ma) that fissions to $^{131-136}$Xe yields a similar age of ~80 Ma (Kunz et al. 1998) since this age range is mostly determined by the decay constant of $^{129}$I. However, if xenon has been escaping for a long period of time (Pujol et al. 2011), then the closure ages have to be corrected for such loss and yield < 50 Ma (range 35-50 Ma) for the "age" of the atmosphere and therefore for the main episodes of volatile delivery. Such timing is more consistent with the early age of lunar formation initially proposed on the basis of W difference between chondrites and Earth. It may represent the time when the Earth retained quantitatively its most volatile elements.

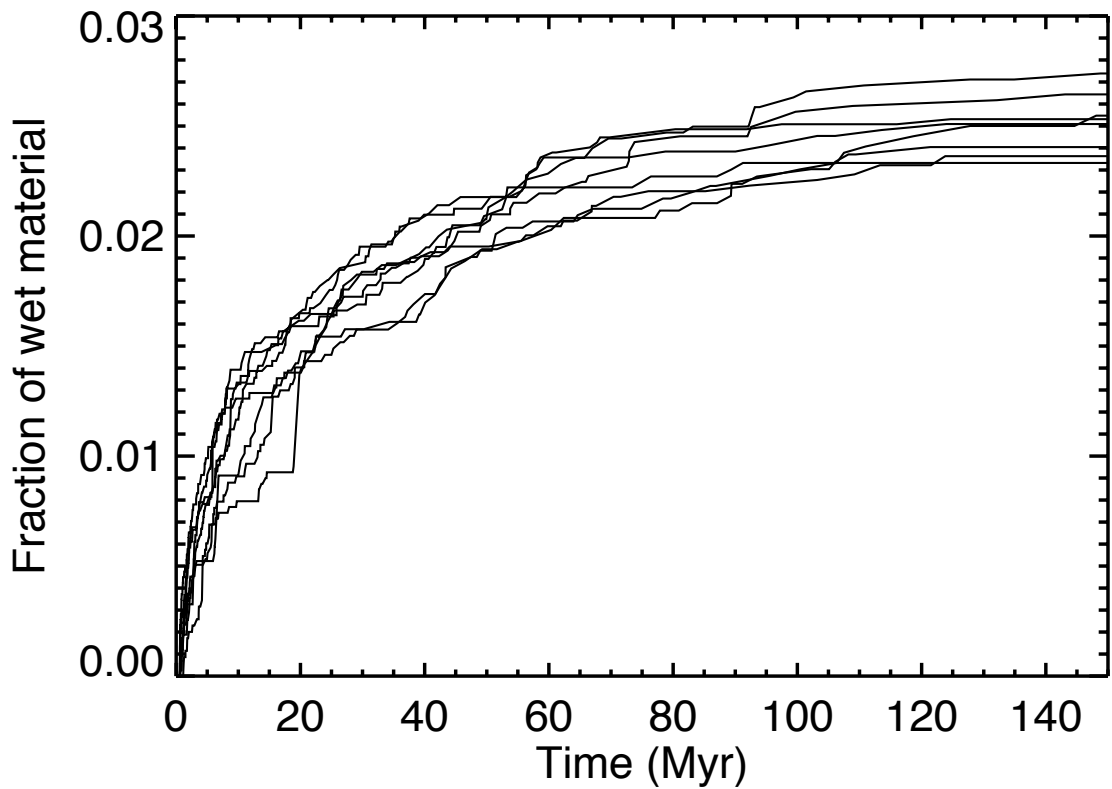

Fig 5: Volatile delivery to Earth analogues in simulations of terrestrial planet formation in the context of the Grand Tack model. Earth analogues are defined here simply as simulated planets with masses of 0.7 to 1.3 Earth masses and orbital semimajor axes between 0.8 and 1.25. The plot shows the fraction of "wet" (carbon-rich C-complex) material accreted onto each simulated planet as a function of time. Each Earth analogue accreted 2-3% of C-complex material in the form of planetesimals that were scattered inward during Jupiter and Saturn's early outward migration but that were accreted on a much longer timescale of tens to 100 Ma. From O'Brien et al, in preparation.

In the context of the Grand Tack model (Walsh et al. 2011) the accretion of volatile-rich (C-complex) material occurs mostly in the first ~50 Ma of the Earth's history (Fig. 5). In dynamical simulations, Earth analogues accrete a total of 2-3% of carbon-rich material, mainly from small planetesimals, with a tail that extends beyond 100 Ma (O'Brien et al, in preparation). In contrast, Earth analogues in simulations with fixed giant planet orbits (e.g., the one in Fig 2) tend to accrete far more C-complex material, typically 5-20% (Raymond et al. 2007; 2009).

Volatile delivery is not perfectly efficient. Some water is lost during accretion due to impact heating for very high-velocity impacts such as those from comets (e.g., Svetsov 2007). Giant impacts may also strip a fraction of water from the growing Earth (Canup & Pierazzo 2006), although the entire atmosphere must be completely removed by an impact before a large fraction of Earth's water can be stripped (Genda & Abe 2005). However, the question of volatile retention merits additional study as only a fraction of the relevant circumstances have been carefully explored.

## CARBON TRAPPING IN EARTH

The Earth experienced very high temperatures during accretion, which might have peaked with the Earth-Moon forming event (Cameron 1997; Pahlevan and Stevenson 2007). Under these conditions, the retention of volatile elements may have been problematic. Noble gases are inert by nature and are presently concentrated in the atmosphere. Chemically reactive volatiles, such as carbon, nitrogen and water, are able to change their speciation and solubilities depending on redox conditions: at low $f_{O_2}$ (below IW) both carbon, nitrogen and hydrogen solubilities in silicates increase drastically (e.g., Kadik et al. 2011), suggesting that significant amounts of these elements could have been retained in a magma ocean under reducing conditions.

From dynamical simulations, (this chapter), contributions from wet planetesimals akin to carbonaceous chondrites are a natural consequence of the evolution of the inner solar system. Such contributions are not likely to have occurred before the late phases of the gaseous protoplanetary disk phase, and more likely towards the last stages of terrestrial accretion, starting after a few Ma and lasting for several tens of Ma. At that time, the size of the Earth was likely to have exceeded half of its present size, and infalling wet planetesimals were quantitatively degassing upon impact (Lange and Ahrens 1982, 1986) generating a $CO_2$-rich steam atmosphere and probably local episodes of intense magmatism. Because the solubility of $CO_2$ is low in molten silicates, only a small fraction of impact generated $CO_2$ could have been trapped in silicates by equilibrium dissolution, and other means of sequestering carbon in the mantle need to be found. At this stage of Earth formation, it is unlikely that reducing conditions below IW could have persisted for long intervals of time, particularly because infalling bodies were oxidized, and the placing of significant amounts of carbon into reduced phases, such as graphite or diamond, was unlikely.

The problem of trapping $CO_2$ from a steam atmosphere into the Earth has been addressed by Sleep et al. (2001). After the Moon-forming impact, the Earth's surface evolved towards a $CO_2$-rich steam atmosphere overlying liquid water in a geologically short period of time (few Ma) due to the bombardment by remnant planetesimals both wet and dry (see Fig 5). An upper limit for the $CO_2$ partial pressure is taken to be the one corresponding to all terrestrial carbon being in the atmosphere in the form of $CO_2$. The atmospheric pressure of $CO_2$ corresponding to a carbon inventory of $2.5 \times 10^{22}$ moles (Zhang and Zindler 1993) is 215 bars, or ~2 kbars if one uses the total bulk Earth carbon content estimated by Marty (2012; $2.6 \times 10^{23}$ moles). The actual $CO_2$ partial pressure could have been much less. Impacts of volatile-rich bodies could have continued for $10^7$-$10^8$ yrs, while transfer of $CO_2$ from the atmosphere and into the mantle may have occurred on a time scale of ~1 Ma or less, corresponding to the duration of a magma ocean episode induced by a large impact (Sleep et al. 2001). For illustration, if impactors all had sizes of 100 km, a typical size for primordial asteroids, ~20,000 of them with 3-4 % carbon would have supplied the required 2 wt.% of Earth's mass to supply terrestrial volatiles. Over 30 Ma, this would have led to an average $CO_2$ partial pressure of only about 100 bars. In these conditions, most water would have been liquid, with a $H_2O$ water pressure was only a few bars.

Based on analogies with present-day mid-ocean ridge systems, Sleep et al. (2001) have argued that atmospheric $CO_2$ precipitated as carbonates in the basaltic crust through hydrothermal circulation. Because of the hotter thermal regime of the Earth, convection was much faster and the magmas were less sluggish, resulting in efficient hydrothermal trapping of carbon in the newly created crust. Foundering of crustal blocks would then transfer carbon into the mantle. Part of "subducted" carbon would then be stored at great depth, escaping recirculation to the surface and degassing. It is possible that, due to the pressure dependence of oxygen fugacity, such recycled carbon could then have been sequestrated in high pressure phases. A critical unknown for this model is the efficiency of transfer into the mantle since a hotter regime may also favour degassing of foundering blocks, by analogy with modern hot subduction zones. Another crucial question is whether the subducted carbon could escape shallow convection and be transferred deeper into the mantle. According to Sleep et al. (2001), $CO_2$ trapping in the oceanic crust and transfer to the mantle would have to have been efficient in order for the temperature at the Earth's surface to allow liquid water to be stable in on the order of $10^6$ to $10^8$ yr, allowing perhaps the development of a biological activity in the 50-100°C temperature range.

There are several areas of research that need to be explored, such as the speciation of carbonates during their transfer into the mantle and the possible existence of carbon-rich fluids in the mantle lasting from this epoch, the direct injection at mantle depth of carbon-rich material during impacts, and the fact that all carbon present in the Earth was probably not at the surface at the same time. We are only opening the magic book of the Earth's formation and early evolution.

*Acknowledgements*

This study is funded by the European Research Council under the European Community's Seventh Framework Programme (FP7/2007-2013 Grant Agreement no. [267255] to BM). SNR thanks the CNRS's PNP program, the Conseil Regional d'Aquitaine, and the Virtual Planetary Laboratory lead team of the NASA Astrobiology Institute. CA was partially funded by the NASA Astrobiology Institute and NASA Cosmochemistry Grant NNX11AG67G. We thank Sara Russell, Adrian Jones and Matthew Genge for comments and suggestions. CRPG contribution n° 2198.

# REFERENCES

Aikawa Y, van Zadelhoff GJ, van Dishoeck EF, Herbst E (2002) Warm molecular layers in protoplanetary disks. Astron. Astrophys 386: 622-632

Aléon J, Engrand C, Robert F, Chaussidon M (2000) Clues on the origin of interplanetary dust particles from the isotopic study of their hydrogen-bearing phases. Geochim Cosmochim Acta 65: 4399-4412

Aléon J, Robert F, Chaussidon M, Marty B (2003) Nitrogen isotopic composition of macromolecular organic matter in interplanetary dust particles. Geochim Cosmochim Acta 67: 3773-3783

Alexander CMO'D et al. (2012) The provenances of asteroids, and their contributions to the volatile inventories of the terrestrial planets. Science 337: 721-723

Alexander CMO'D, Fogel M, Yabuta H, Cody GD (2007) The origin and evolution of chondrites recorded in the elemental and isotopic compositions of their macromolecular organic matter. Geochim Cosmochim Acta 71: 4380-4403

Alexander CMO'D (2005) Re-examining the role of chondrules in producing the volatile element fractionations in chondrites. Meteoritics Planet Sci 40: 943-965

Alexander CMO'D et al. (2010) Deuterium enrichments in chondritic macromolecular material – Implications for the origin, evolution of organics, water and asteroids. Geochim Cosmochim Acta 74: 4417-4437

Alexander CMO'D et al. (1998) The origin of chondritic macromolecular organic matter: A carbon and nitrogen isotope study. Meteoritics Planet Sci 33: 603-622

Allègre CJ, Hofmann AW, O'Nions RK (1996) The argon constraints on mantle structure. Geophys Res Lett 23: 3555-3557

Anders E, Grevesse N (1989) Abundances of the elements : Meteoritic and Solar. Geochim Cosmochim Acta 53: 197-214

Arevalo R, McDonough WF, Luong M (2009) The K/U ratio of the silicate Earth: Insights into mantle composition, structure and thermal evolution. Earth Planet Sci Lett 278: 361-369

Arpigny C et al. (2003) Anomalous nitrogen isotope ratio in comets. Science 301: 1522-1524

Asplund M, Grevesse N, Sauval AJ, Scott P (2009) The Chemical Composition of the Sun. Ann Rev Astron Astrophys 47: 481-522

Benz, W., Slattery, W.L. and Cameron, A.G.W., 1986. The origin of the Moon and the single impact hypothesis .1. Icarus, 66: 515-535

Bianchi D et al. (2010) Low helium flux from the mantle inferred from simulations of oceanic helium isotope data. Earth Planet Sci Lett 297: 379-386

Bland PA et al. (1996) The flux of meteorites to the Earth over the last 50 000 years. Mon. Not. R. Astron. Soc 283: 551-565

Blum J, Wurm G (2008) The growth mechanisms of macroscopic bodies in protoplanetary disks. Ann Rev Astron Astrophys 46: 21-56

Bockelée-Morvan D et al. (2008) Large excess of heavy nitrogen in both hydrogen cyanide and cyanogen from comet 17P/Holmes. Astrophys J 679 : L49-L52

Bockelée-Morvan D, Crovisier J, Mumma MJ, Weaver HA (2004) The composition of cometary volatiles. Festou MC, Keller U, Weaver HA (eds) Comets 2. Univ. Arizona Press: 391-423

Bockelée-Morvan D et al. (1998) Deuterated water in comet C 1996 B2 (Hyakutake) and its implications for the origin of comets. Icarus 133: 147-162

4
Bodmer P, Bochsler P (1998) The helium isotopic ratio in the solar wind and ion fractionation in the corona by inefficient Coulomb drag. Astron Astrophys 337: 921-927

Boss AP (1998) Temperatures in protoplanetary disks. Ann Rev Earth Planet Sci 26: 53-80

Boss, A. P. (1997) Giant planet formation by gravitational instability. Science 276: 1836-1839

Bottke WF et al. (2012) An Archaean heavy bombardment from a destabilized extension of the asteroid belt. Nature 485: 78-81

Bradley JP (2003) Interplanetary dust particles. Davis AM (ed), Meteorites, Comets and Planets. Treatise on geochemistry. Elsevier-Pergamon, Oxford: 689-712

Brearley AJ and Jones RH (1998) Chondritic meteorites. Papike JJ (ed), Planetary Materials. Mineralogical Society of America, Washington DC: 3:1-3:398

Brownlee D et al. (2006) Comet 81P/Wild 2 under a microscope. Science 314: 1711-1716

Burbidge EM et al. (1957) Synthesis of the elements in stars. Rev Modern Phys, 29: 547-650

Burkhardt C et al. (2011) Molybdenum isotope anomalies in meteorites: Constraints on solar nebula evolution and origin of the Earth. Earth Planet Sci Lett, 312: 390-400

Busemann H et al. (2006) Interstellar chemistry recorded in organic matter from primitive meteorites. Science 312: 727-730

Busemann H et al. (2009) Ultra-primitive interplanetary dust particles from the comet 26P/Grigg-Skjellerup dust stream collection. Earth Planet Sci Lett 288: 44-57

Cameron AGW (1997) The origin of the moon and the single impact hypothesis .5. Icarus 126: 126-137

Canup RM, Asphaug E (2001) Origin of the Moon in a giant impact near the end of the Earth's formation. Nature 412: 708-712

Canup RM, Pierazzo E (2006) Retention of Water During Planet-Scale Collisions. 37th Annual Lunar and Planetary Science Conference 37: 2146-

Caro G, Bourdon B, Halliday AN, Quitté G. (2008) Super-chondritic Sm/Nd ratios in Mars, the Earth and the Moon. Nature 452: 336-339

Cartigny P, Harris JW, Javoy M (1998) Subduction-related diamonds? - The evidence for a mantle-derived origin from coupled $\delta^{13}C$-$\delta^{15}N$ determinations. Chem Geol 147: 147-159

Cartigny, P., Harris, J.W., Javoy M (2001) Diamond genesis, mantle fractionations and mantle nitrogen content: a study of $\delta^{13}C$-N concentrations in diamonds. Earth Planet Sci Lett 185: 85-98

Cartigny P, Pineau F, Aubaud C, Javoy M (2008) Towards a consistent mantle carbon flux estimate: Insights from volatile systematics ($H_2O$/Ce, $\delta D$, $CO_2$/Nb) in the North Atlantic mantle (14 degrees N and 34 degrees N). Earth Planet Sci Lett 265: 672-685

Chambers JE (2010) Planetesimal formation by turbulent concentration. Icarus 208: 505-517

Chiang E, Youdin AN (2010) Forming planetesimals in solar and extrasolar nebulae. Ann Rev Earth Planet Sci 38: 493-522

Ciesla FJ, Cuzzi JN (2006) The evolution of the water distribution in a viscous protoplanetary disk. Icarus 181: 178-204

Cody GD, Alexander CMO'D, Tera F (2002) Solid state ($^1H$ and $^{13}C$) NMR spectroscopy of the insoluble organic residue in the Murchison meteorite: A self-consistent quantitative analysis. Geochim Cosmochim Acta 66 : 1851-1865

Craig H, Clarke WB, Beg MA (1975) Excess $^3He$ in deep waters on the East Pacific Rise. Earth Planet Sci Lett 26: 125-132

Cuzzi JN, Hogan RC, Shariff K (2008) Toward planetesimals: Dense chondrule clumps in the protoplanetary nebula. Astrophys J 687: 1432-1447

Dasgupta R, Walker D (2008) Carbon solubility in core melts in a shallow magma ocean environment and distribution of carbon between the Earth's core and the mantle. Geochim Cosmochim Acta 72: 4627-4641

Dauphas N (2003) The dual origin of the terrestrial atmosphere. Icarus 165: 326-339

Dauphas N, Pourmand A (2011) Hf-W-Th evidence for rapid growth of Mars and its status as a planetary embryo. Nature 473: 489-492

Dauphas N, Davis AM, Marty B, Reisberg L (2004) The cosmic molybdenum-ruthenium isotope correlation. Earth Planet Sci Lett 226: 465-475

De Gregorio BT et al. (2010) Isotopic anomalies in organic nanoglobules from Comet 81P/Wild 2: Comparison to Murchison nanoglobules and isotopic anomalies induced in terrestrial organics by electron irradiation. Geochim Cosmochim Acta 74: 4454-4470

Deines P (1980) The carbon isotopic composition of diamonds - relationship to diamond shape, color, occurrence and vapor composition. Geochim Cosmochim Acta 44: 943-961

Delsemme AH (1991) Nature and history of the organic compounds in comets - an astrophysical view. Newburn Jr RL, Neugebauer M, Rahe J (eds), Comets in the post-Halley era. Kluwer Academic Press, Dordrecht: 377-428

Des Marais DJ, Moore JG (1984) Carbon and its isotopes in mid-oceanic basaltic glasses. Earth Planet Sci Lett 62: 43-57

Dodson-Robinson SE, Willacy K, Bodenheimer P, Turner NJ, Beichman CA (2009) Ice lines, planetesimal composition and solid surface density in the solar nebula. Icarus 200: 672-693

Draine BT, Li A (2007) Infrared emission from interstellar dust. IV. The silicate-graphite-PAH model in the post-Spitzer era. Astrophys J 657: 810-837

Duprat J et al. (2010) Extreme deuterium excesses in ultracarbonaceous micrometeorites from central Antarctic snow. Science 328: 742-745

Engrand C, Maurette M (1998) Carbonaceous micrometeorites from Antartica. Meteorit Planet Sci 33: 565-580

Engrand C et al. (1999) Extraterrestrial water in micrometeorites and cosmic spherules from Antarctica: an ion microprobe study. Meteorit Planet Sci 34: 773-786

Fernandez JA, Ip WH (1984) Some dynamical aspects of the accretion of Uranus and Neptune - the exchange of orbital angular-momentum with planetesimals. Icarus 58: 109-120

Floss C et al. (2006) Identification of isotopically primitive interplanetary dust particles: A NanoSIMS isotopic imaging study. Geochim Cosmochim Acta 70: 2371-2399

Floss C, Stadermann FJ (2009) High abundances of circumstellar and interstellar C-anomalous phases in the primitive CR3 chondrites QUE 99177 and MET 00426. Astrophys J 697: 1242-1255

Flynn GJ et al. (2006) Elemental compositions of comet 81P/Wild 2 samples collected by Stardust. Science 314: 1731-1735

Garaud P, Lin DNC (2007) The effect of internal dissipation and surface irradiation on the structure of disks, and the location of the snow line around Sun-like stars. Astrophys J 654: 606-624.

Garvie LAJ, Buseck PR (2006) Carbonaceous materials in the acid residue from the Orgueil carbonaceous chondrite meteorite. Meteoritics Planet Sci 41: 633-642

Geiss J, Bochsler P (1982) Nitrogen isotopes in the solar system. Geochim Cosmochim Acta 46: 529-548

Geiss J, Gloeckler G (2003) Isotopic composition of H, He and Ne in the protosolar cloud. Space Sci Rev 106: 3-18

Genda H, Abe Y (2005) Enhanced atmospheric loss on protoplanets at the giant impact phase in the presence of oceans.  Nature 433:  842-844

Gilmour I (2003) Structural and isotopic analysis of organic matter in carbonaceous chondrites. Davis AM (ed), Meteorites, Comets, and Planets. Treatise on geochemistry. Elsevier-Pergamon, Oxford: 269-290

Goldreich P,  Tremaine S (1980) Disk-satellite interactions. Astrophys J 241: 425-441

Gomes R, Levison HF, Tsiganis K,  Morbidelli A (2005) Origin of the cataclysmic Late Heavy Bombardment period of the terrestrial planets. Nature 435: 466-469

Greenberg R, Wacker JF, Hartmann WK,  Chapman CR (1978) Planetesimals to planets - numerical simulation of collisional evolution. Icarus  35: 1-26

Guillot T (2005) The interiors of giant planets: Models, and outstanding questions. Ann Rev Earth Planet Sci 33: 493-530

Gourier D et al. (2008) Extreme deuterium enrichment of organic radicals in the Orgueil meteorite: Revisiting the interstellar interpretation? Geochim Cosmochim Acta 72: 1914-1923

Grady MM, Swart PK and Pillinger CT (1982) The variable carbon isotopic composition of type 3 ordinary chondrites. J Geophys Res 87: A289-A296

Grady MM, Wright IP, Carr LP, Pillinger CT (1986) Compositional differences in enstatite chondrites based on carbon and nitrogen stable isotope measurements. Geochim Cosmochim Acta 50: 2799-2813

Grady MM, Wright IP, Swart PK,  Pillinger CT (1988) The carbon and oxygen isotopic composition of meteoritic carbonates. Geochim Cosmochim Acta 52: 2855-2866

Grady MM, Verchovsky AB, Franchi IA, Wright IP and Pillinger CT (2002) Light element geochemistry of the Tagish Lake CI2 chondrite: Comparison with CI1 and CM2 meteorites. Meteoritics Planet Sci 37: 713-735

Greenberg JM (1998) Making a comet nucleus. Astron Astrophys 330: 375-380

Haisch KE, Lada EA, Pina RK, Telesco CM,  Lada CJ (2001) A mid-infrared study of the young stellar population in the NGC 2024 cluster. Astron J 121: 1512-1521

Hansen BMS (2009) Formation of the terrestrial planets from a narrow annulus. Astrophys J 703: 1131-1140

Hartogh P et al. (2011) Ocean-like water in the Jupiter-family comet 103P/Hartley 2. Nature 478: 218-220

Hashizume K, Chaussidon M, Marty B,  Terada K (2004) Protosolar carbon isotopic composition: Implications for the origin of meteoritic organics. Astrophys J 600: 480-484

Hayes JF,  Waldbauer JR (2006) The carbon cycle and associated redox processes through time. Phil Trans Royal Soc London, B361: 931-950

Herbst E, van Dishoeck EF (2009) Complex organic interstellar molecules, Ann Rev Astron Astrophys 47: 427-480

Herd CDK et al. (2011) Origin and evolution of prebiotic organic matter as inferred from the Tagish Lake meteorite. Science 332: 1304-1307

Hirschmann MM,  Dasgupta R (2009) The H/C ratios of Earth's near-surface and deep reservoirs, and  consequences for deep Earth volatile cycles. Chem Geol  262: 4-16

Holser WT, Schidlowski M, MacKenzie FT, Maynard JB (1988) Geochemical cycles of carbon, and sulfur. Chemical cycles in the evolution of Earth. Wiley, New York: 105-173

Horn B, Lyra W, Mac Low M-M, and Sandor Z (2012) Orbital Migration of Interacting Low-mass Planets in Evolutionary Radiative Turbulent Models. Astrophys J 750: 34-42

Howard AW et al. (2010) The California planet survey. I. Four new giant exoplanets. Astrophys J 721: 1467-1481

Hubickyj O, Bodenheimer P, Lissauer JJ (2005) Accretion of the gaseous envelope of Jupiter around a 5-10 Earth-mass core. Icarus 179: 415-431

Hughes ALH, Armitage PJ (2010) particle transport in evolving protoplanetary disks: Implications for results from Stardust. Astrophys J 719: 1633-1653

Huss GR (1990) Ubiquitous interstellar diamond and SiC in primitive chondrites: abundances reflect metamorphism. Nature 347: 159-162

Huss GR, Lewis RS (1994) Noble gases in presolar diamonds. II. Component abundances reflect thermal processing. Meteoritics 29: 811-829

Inaba S, Ikoma M (2003) Enhanced collisional growth of a protoplanet that has an atmosphere. Astron Astrophys 410: 711-723

Ida S, Makino JI (1992) N-body simulation of gravitational interaction between planetesimals and a protoplanet .1. Velocity distribution of planetseimals. Icarus 96: 107-120

Ikoma M, Nakazawa K, Emori H (2000) Formation of giant planets: Dependences on core accretion rate and grain opacity. Astrophys J 537: 1013-1025

Javoy M, Pineau F, Allègre CJ (1982) Carbon geodynamic cycle. Nature 300: 171-173

Javoy M, Pineau F, Delorme H (1986) Carbon and nitrogen isotopes in the mantle. Chem Geol 57: 41-62

Jessberger EK, Christoforidis A, Kissel J (1988) Aspects of the major element composition of Halley's dust. Nature 332: 691-695

Johansen A, Youdin A (2007) Protoplanetary disk turbulence driven by the streaming instability: Nonlinear saturation and particle concentration. Astrophys J 662: 627-641

Kadik AA et al. (2011) Influence of oxygen fugacity on the solubility of nitrogen, carbon and hydrogen in $FeO-Na_2O-SiO_2-Al_2O_3$ melts in equilibrium with metallic iron at 1.5 GPa and 1400 degrees C. Geochem. Int 49: 429-438

Kehm K, Flynn GJ, Sutton SR, Hohenberg CM (2002) Combined noble gas and trace element measurements on individual stratospheric interplanetary dust particles. Meteoritics Planet Sci 37: 1323-1335

Keller LP et al. (2004) The nature of molecular cloud material in interplanetary dust. Geochim Cosmochim Acta 68: 2577-2589

Kerridge F (1985) Carbon, hydrogen and nitrogen in carbonaceous chondrites: abundances, and isotopic compositions in bulk samples. Geochim Cosmochim Acta 49: 1707-1714

Kenyon SJ, Bromley BC (2006) Terrestrial planet formation. I. The transition from oligarchic growth to chaotic growth. Astron J 131: 1837-1850

Kimura K, Lewis RS, Anders E (1974) Distribution of gold and rhenium between nickel-iron and silicate melts - Implications for abundance of siderophile elements on Earth and Moon. Geochim Cosmochim Acta 38: 683-701

Kirsh DR, Duncan M, Brasser R, Levison HF (2009) Simulations of planet migration driven by planetesimal scattering. Icarus 199: 197-209

Kissel J, Krueger FR (1987) The organic component in dust from comet Halley as measured by the PUMA mass spectrometer on board Vega 1. Nature 326: 755-760

Kleine T, Munker C, Mezger K, Palme H (2002) Rapid accretion and early core formation on asteroids and the terrestrial planets from Hf-W chronometry. Nature, 418: 952-955

Kobayashi C, Nakasato N (2011) Chemodynamical simultions of the Milky Way galaxy. Astrophys J 729: 16 DOI: 10.1088/0004-637X/729/1/16

Kokubo E, Ida S (1998) Oligarchic growth of protoplanets. Icarus 131: 171-178

Kokubo E, Ida S (2002) Formation of protoplanet systems and diversity of planetary systems. Astrophys J 581: 666-680

Kress ME, Tielens A, Frenklach M (2010) The 'soot line': Destruction of presolar polycyclic aromatic hydrocarbons in the terrestrial planet-forming region of disks. Adv Space Res 46: 44-49

Kunz J, Staudacher T, Allègre CJ (1998) Plutonium-fission xenon found in the Earth's mantle. Science 280: 877-880

Lambrechts M, Johansen A (2012) Rapid growth of gas-giant cores by pebble accretion. Astron. & Astrophys. 544: A32

Lange MA, Ahrens TJ (1982) The evolution of an impact-generated atmosphere. Icarus 51: 96-120.

Lange MA, Ahrens TJ (1986) Shock-induced $CO_2$ loss from $CaCO_3$ - Implications for early planetary atmospheres. Earth Planet Sci Lett 77: 409-418

Laskar J, Gastineau M (2009) Existence of collisional trajectories of Mercury, Mars and Venus with the Earth. Nature 459: 817-819

Lécuyer, C, Gillet P, Robert F (1998) The hydrogen isotope composition of seawater and the global water cycle. Chem Geol 145: 249-261

Leinhardt ZM, Richardson DC, Lufkin G, Haseltine J (2009) Planetesimals to protoplanets - II. Effect of debris on terrestrial planet formation. Monthly Notices of the Royal Astronomical Society 396: 718-728

Levison HF, Thommes E, Duncan MJ (2010) Modeling the formation of giant planet cores : I. Evaluating key processes. Astron J 139: 1297-1314

Lin DNC, Papaloizou J (1986) On the tidal interaction between protoplanets and the protoplanetary disk. 3. Orbital migrations of protoplanets. Astrophys J 309: 846-857

Lissauer JJ, Hubickyj O, D'Angelo G, Bodenheimer P (2009) Models of Jupiter's growth incorporating thermal and hydrodynamic constraints. Icarus 199: 338-350.

Lodders K (2003) Solar system abundances and condensation temperatures of the elements. Astrophys J 591: 1220-1247

Lodders K (2004) Jupiter formed with more tar than ice. Astrophys J 611: 587-597

Love S, Brownlee DE (1993) A direct measurment of the terrestrial mass accretion rate of cosmic dust. Science 262: 550-553

Lynden-Bell D, Pringle JE (1974) Evolution of viscous disks and origin of nebular variables. Month Not Royal Astron Soc 168: 603-637

Lyra W, Paardekooper S-J, and Mac Low, M-M (2010) Orbital Migration of Low-mass Planets in Evolutionary Radiative Models: Avoiding Catastrophic Infall. The Astrophysical Journal 715: L68-L73.

Manfroid J et al. (2009) The CN isotopic ratios in comets. Astron Astrophys 503: 613-U354.

Marty B (2012) The origins and concentrations of water, carbon, nitrogen and noble gases on Earth. Earth Planet Sci Lett 313-314: 56-66

Marty B, Chaussidon M, Wiens RC, Jurewicz AJG, Burnett DS (2011) A $^{15}$N-poor isotopic composition for the solar system as shown by Genesis solar wind samples. Science 332: 1533-1536

Marty B, Jambon A (1987) $C/^3He$ in volatile fluxes from the solid Earth : Implications for carbon geodynamics. Earth Planet Sci Lett 83: 16-26

Marty B et al. (2008) Helium and neon abundances and compositions in cometary matter. Science 319: 75-78

Marty B, Tolstikhin IN (1998) CO$_2$ fluxes from mid-ocean ridges, arcs and plumes. Chem Geol 145: 233-248
Marty B, Zimmermann L (1999) Volatiles (He, C, N, Ar) in mid-ocean ridge basalts: Assessment of shallow-level fractionation and characterization of source composition. Geochim Cosmochim Acta 63: 3619-3633
Marty B, Robert P, Zimmermann L (2005) Nitrogen and noble gases in micrometeorites. Meteorit Planet Sci 40: 881-894
Masset F, Snellgrove M (2001) Reversing type II migration: resonance trapping of a lighter giant protoplanet. Month Not Royal Astron Soc 320: L55-L59
Matrajt G et al. (2003) A nuclear microprobe study of the distribution and concentration of carbon and nitrogen in Murchison and Tagish Lake meteorites, Antarctic micrometeorites, and IDPs : Implications for astrobiology. Meteorit Planet Sci 38: 1585-1600
Matrajt G et al. (2008) Carbon investigation of two Stardust particles: A TEM, NanoSIMS, and XANES study. Meteoritics Planet Sci 43: 315-334
Matrajt G, Messenger S, Brownlee D, Joswiak D (2012) Diverse forms of primordial organic matter identified in interplanetary dust particles. Meteoritics Planet Sci 47: 525-549
McKeegan KD et al. (2006) Isotopic compositions of cometary matter returned by Stardust. Science 314: 1724-1728
Meech KJ et al. (2011) EPOXI; Comet 103P/Hartley 2 observations from a worldwide campaign. Astrophys J 734: 9-16
Messenger S (2000) Identification of molecular-cloud material in interplanetary dust particles. Nature 404: 968-971
Messenger S et al. (2003a) Samples of stars beyond the solar system: Silicate grains in interplanetary dust. Science 300: 105-108
Messenger S, Stadermann FJ Floss C, Nittler LR, Mukhopadhyay S (2003b) Isotopic signatures of presolar materials in interplanetary dust. Space Sci Rev 106: 155-172
Meyer BS, Zinner E (2006) Nucleosynthesis. Lauretta DS, H.Y. McSween Jr HY (eds), Meteorites and the Early Solar System II. Univ. Arizona Press, Tucson: 69-108
Meyer MR et al. (2008) Evolution of mid-infrared excess around sun-like stars: constraints on models of terrestrial planet formation. Astrophys J 673: L181-L184
Nieva MF et al. (2011) Fundamental parameters of "normal" B stars in the solar neighborhood. Neiner, C (ed) Active OB stars: Structure, evolution, mass-loss, and critical limits. IAU Symposium Proceedings Series. Cambridge Univ Press, Cambridge: 566-570
Nittler LR (2003) Presolar stardust in meteorites: recent advances and scientific frontiers. Earth Planet Sci Lett 209: 259-273
Mizuno H (1980) Formation of the giant planets. Progr Theor Phys 64: 544-557
Morbidelli A et al. (2000) Source regions and timescales for the delivery of water to the Earth. Meteoritics Planet Sci 35: 1309-1320
Morbidelli A, Crida A (2007) The dynamics of Jupiter and Saturn in the gaseous protoplanetary disk. Icarus 191: 158-171
Morbidelli A, Tsiganis K, Crida A, Levison HF, Gomes R (2007) Dynamics of the giant planets of the solar system in the gaseous protoplanetary disk and their relationship to the current orbital architecture. Astron J 134: 1790-1798
Morbidelli A, Bottke WF, Nesvorny D, Levison HF (2009) Asteroids were born big. Icarus 204: 558-573

Morbidelli A, Lunine JI, O'Brien DP, Raymond SN, Walsh KJ (2012) Building terrestrial planets. Ann Rev Earth Planet Sci 40, 251-275

Morishima R, Schmidt MW, Stadel J, Moore B (2008) Formation and accretion history of terrestrial planets from runaway growth through to late time: Implications for orbital eccentricity. Astrophys J 685: 1247-1261

Morishima R, Stadel J, Moore B (2010) From planetesimals to terrestrial planets: N-body simulations including the effects of nebular gas and giant planets. Icarus 207: 517-535

Mumma MJ, Charnley SB (2012) The chemical composition of comets - Emerging taxonomies and natal heritage. Ann Rev Astron Astrophys 49: 471-524

Muralidharan K et al. (2009) Water in the inner solar system: Insights from atomistic and electronic-structure calculations. Meteoritics Planet Sci 44: A136

O'Brien DP, Morbidelli A, Levison HF (2006) Terrestrial planet formation with strong dynamical friction. Icarus 184: 39-58

O'Brien DP, Walsh KJ, Raymond SN, Morbidelli A, Mandell AM (2012) Terrestrial Planet Formation in the 'Grand Tack' Scenario. In preparation

Ott U (2002) Noble gases in meteorites - Trapped components. Rev Mineral Geochem 47: 71-100

Ott U et al. (2012) New attempts to understand nanodiamond stardust. Publ Astron Soc Aust 29: 90-97

Otting W, Zähringer J (1967) Total carbon content and primordial rare gases in chondrites. Geochim Cosmochim Acta 31: 1949-1956

Owen T, Bar-Nun A, Kleinfeld I (1992) Possible cometary origin of heavy noble gases in the atmospheres of Venus, Earth and Mars. Nature 358: 43-46

Ozima M, Podosek FA (2002) Noble gas geochemistry. Cambridge University Press, Cambridge, 286 pp.

Pahlevan K, Stevenson DJ (2007) Equilibration in the aftermath of the lunar-forming giant impact. Earth Planet Sci Lett 262: 438-449

Paardekooper S-J, Baruteau C, and Kley W (2011) A torque formula for non-isothermal Type I planetary migration - II. Effects of diffusion. Month Not Royal Astron Soc 410: 293-303.

Papaloizou JCB, Terquem C (2006) Planet formation and migration. Rep Progress Phys: 119-180

Pepin RO (1991) On the origin and early evolution of terrestrial planet atmospheres and meteoritic volatiles. Icarus 92: 2-79

Pepin RO (2006) Atmospheres on the terrestrial planets: Clues to origin and evolution. Earth Planet Sci Lett 252: 1-14

Pierens A, Nelson RP (2008) Constraints on resonant-trapping for two planets embedded in a protoplanetary disc. Astron Astrophys 482: 333-340

Pierens A, Raymond SN (2011) Two phase, inward-then-outward migration of Jupiter and Saturn in the gaseous solar nebula. Astron Astrophys 533: 14

Pizzarello S, Cooper GW, Flynn GJ (2006) The nature and distribution of the organic material in carbonaceous chondrites and interplanetary dust particles. Lauretta DS, McSween Jr HY (eds), Meteorites and the Early Solar System II. University of Arizona Press, Tucson: 625-651

Pollack JB et al. (1994) Composition and radiative properties of grains of molecular clouds and accretion disks. Astrophys J 421: 615-639

Pujol M, Marty B, Burgess R (2011) Chondritic-like xenon trapped in Archean rocks: A possible signature of the ancient atmosphere. Earth Planet Sci Lett 308: 298-306

Raymond SN, Armitage PJ, Gorelick N (2010) Planet-planet scattering in planetesimal disks. II. Predictions for outer extrasolar planetary systems. Astrophys J 711: 772-795

Raymond SN, Quinn T, Lunine JI 2004. Making other earths: dynamical simulations of terrestrial planet formation and water delivery. Icarus 168: 1-17

Raymond SN, Quinn T, Lunine JI 2006) High-resolution simulations of the final assembly of Earth-like planets I. Terrestrial accretion and dynamics. Icarus 183: 265-282

Raymond SN, Quinn T, Lunine JI (2007) High-resolution simulations of the final assembly of Earth-like planets. 2. Water delivery and planetary habitability. Astrobiology, 7(1): 66-84

Raymond SN, O'Brien DP, Morbidelli A, Kaib NA (2009) Building the terrestrial planets: Constrained accretion in the inner Solar System. Icarus 203: 644-662

Raymond SN et al. (2011) Debris disks as signposts of terrestrial planet formation. Astron Astrophys 530: 23

Robert F (2003) The D/H ratio in chondrites. Space Sci Rev 106: 87-101

Robert F, Epstein S (1982) The concentration and isotopic composition of hydrogen, carbon and nitrogen in carbonaceous chondrites. Geochim Cosmochim Acta 46: 81-95

Ronov AB, Yaroshevskhiy AA (1976) A new model for the chemical structure of the Earth's crust. Geochem. Int., 13: 89-121

Saal AE, Hauri EH, Langmuir CH, Perfit MR (2002) Vapour undersaturation in primitive mid-ocean-ridge basalt and the volatile content of Earth's upper mantle. Nature 419: 451-455

Safronov VS, Zvjagina EV (1969) Relative sizes of largest bodies during accumulation of planets. Icarus 10: 109-115

Salters VJM, Stracke A, (2004) Composition of the depleted mantle. Geochem Geophys Geosyst 5: Q05B07, doi:10.1029/2003GC000597

Sasselov DD, Lecar M (2000) On the snow line in dusty protoplanetary disks. Astrophys J 528: 995-998

Schramm LS, Brownlee DE, Wheelock MM (1989) Major element composition of stratospheric micrometeorites. Meteoritics, 24: 99-112

Scott ERD, Krot AN (2003) Chondrites and their components. Davis AM (ed), Meteorites, Comets and Planets. Treatise on Geochemistry. Elsevier-Pergamon, Oxford: 143-200

Sharp ZD, Draper DS, Agee CB (2009) Core-mantle partitioning of chlorine and a new estimate for the hydrogen abundance of the Earth. Lunar Planet. Sci. #1209

Sleep NH, Zahnle K, Neuhoff PS (2001) Initiation of clement surface conditions on the earliest Earth. Proc Nat Acad Sci USA 98: 3666-3672

Stevenson DJ, Lunine JI (1988) Rapid formation of Jupiter by diffusive redistribution of water-vapor in the solar nebula. Icarus 75: 146-155

Svetsov VV (2007) Atmospheric erosion and replenishment induced by impacts of cosmic bodies upon the Earth and Mars. Solar System Research 41: 28-41

Thomas KL, Blanford GE, Keller LP, Klöck W, McKay DS (1993) Carbon abundance and silicate mineralogy of anhydrous interplanetary dust particles. Geochim Cosmochim Acta 57: 1551-1566

Thommes EW, Duncan MJ, Levison HF (2003) Oligarchic growth of giant planets. Icarus 161: 431-455

Tielens AGGM (2008) Interstellar polycyclic aromatic hydrocarbon molecules. Ann Rev Astron Astrophys 46: 289-337

Tolstikhin IN, O'Nions RK (1994) The Earth missing xenon - a combination of early degassing and rare gas loss from the atmosphere. Chem Geol 115: 1-6

Touboul M, Kleine T, Bourdon B, Palme H, Wieler R (2007) Late formation and prolonged differentiation of the Moon inferred from W isotopes in lunar metals. Nature 450: 1206-1209

Truran Jr JW, Heger A (2003) Origin of the Elements. Davis AM (ed), Treatise on Geochemistry. Pergamon, Oxford: 1-15

Tsiganis K, Gomes R, Morbidelli A, Levison HF (2005) Origin of the orbital architecture of the giant planets of the Solar System. Nature 435: 459-461

Turcotte S, Richer J, Michaud G, Iglesias CA, Rogers FJ (1998) Consistent solar evolution model including diffusion and radiative acceleration effects. Astrophys J 504: 539-558

Walsh KJ, Morbidelli A, Raymond SN, O'Brien DP, Mandell AM (2005) A low mass for Mars from Jupiter's early gas-driven migration. Nature 475: 206-209

Weidenschilling SJ (2011) Initial sizes of planetesimals and accretion of the asteroids. Icarus 214: 671-684

Wetherill GW (1975) Radiometric chronology of early Solar-System. Ann Rev Nucl Particle Sci 25: 283-328

Wetherill GW (1985) Occurrence of giant impacts during the growth of the terrestrial planets. Science 228: 877-879

Wetherill GW (1990) Formation of the Earth. Ann Rev Earth Planet Sci 18: 205-256

Wetherill GW (1991) Occurrence of Earth-like bodies in planetary systems. Science 253: 535-538

Wetherill GW, Stewart GR (1993) Formation of planetary embryos - effects of fragmentation, low relative velocity, and independent variation of eccentricity and inclination. Icarus 106: 190-209

Wood BJ (1993) Carbon in the core. Earth Planet Sci Lett 117: 593-607

Woolum DS, Cassen P (1999) Astronomical constraints on nebular temperatures: Implications for planetesimal formation. Meteoritics Planet Sci 34: 897-907

Yang J, Epstein S (1983) Interstellar organic-matter in meteorites. Geochim Cosmochim Acta 47: 2199-2216

Yin Q et al. (2002) A short timescale for terrestrial planets from Hf-W chronometry. Nature 418: 949-952

Zhang Y, Zindler A (1993) Distribution and evolution of carbon and nitrogen in Earth. Earth Planet Sci Lett 117: 331-345

Zubko V, Dwek E, Arendt RG (2004.) Interstellar dust models consistent with extinction, emission, and abundance constraints. Astrophys J Suppl Series 152: 211-249